\documentclass[usenatbib]{mn2e}
\usepackage{bm,amssymb}
\usepackage{graphicx}
\usepackage{aas_macros}

\newcommand{\qeos}{q_{\rm eos}}
\newcommand{\kms}{{\rm \, km \, s^{-1}}}

\title[BH Growth with Feedback by Radiation Pressure]
{The Growth of Massive Black Holes in Galaxy Merger Simulations with Feedback by Radiation Pressure}
\author[J. DeBuhr et. al.]{Jackson~DeBuhr,$^{1,2}$ Eliot~Quataert,$^{1,2}$ and Chung-Pei~Ma$^2$ \\
  $^1$Department of Physics, University of California, Berkeley, CA 94720, USA \\
  $^2$Department of Astronomy and Theoretical Astrophysics Center,
  University of California, Berkeley, CA 94720, USA}

\begin{document}

\maketitle

 \begin{abstract}
   We study the growth of massive black holes (BH) in galaxies using
   smoothed particle hydrodynamic simulations of major galaxy mergers
   with new implementations of BH accretion and feedback.  The effect
   of BH accretion on gas in its host galaxy is modeled by depositing
   momentum at a rate $\sim \tau L/c$ into the ambient gas, where $L$
   is the luminosity produced by accretion onto the BH and $\tau$ is
   the wavelength-averaged optical depth of the galactic nucleus to
   the AGN's radiation (a free parameter of our model).  The accretion
   rate onto the BH is relatively independent of our subgrid accretion
   model and is instead determined by the BH's dynamical impact on its
   host galaxy: BH accretion is thus self-regulated rather than
   ``supply limited.''  We show that the {final} BH mass and total
   stellar mass formed during a merger are more robust predictions of
   the simulations than the {time dependence} of the star formation
   rate or BH accretion rate.  In particular, the latter depend on the
   assumed interstellar medium physics, which determines when and
   where the gas fragments to form star clusters; this in turn affects
   the fuel available for further star formation and BH growth.
   Simulations over a factor of $\sim 30$ in galaxy mass are
   consistent with the observed $M_{BH}-\sigma$ relation for a mean
   optical depth of $\tau \sim 25$.  This requires that most BH growth
   occur when the galactic nucleus is optically thick to far-infrared
   radiation, consistent with the hypothesized connection between
   ultra-luminous infrared galaxies and quasars.  We find tentative
   evidence for a shallower $M_{BH}-\sigma$ relation in the lowest
   mass galaxies, $\sigma \lesssim 100 \kms$.  Our results demonstrate
   that feedback-regulated BH growth and consistency with the observed
   $M_{BH}-\sigma$ relation do not require that BH feedback terminate
   star formation in massive galaxies or unbind large quantities of
   cold gas.

\end{abstract}

\section{Introduction}

Feedback from an active galactic nucleus (AGN) has been invoked to
resolve a number of observational problems in galaxy formation: (1) to
explain the tight observed
\citep{ferrarese2000,gebhardt2000,haring2004} correlations between
central black hole (BH) and galaxy properties such as the
$M_{BH}-\sigma$ and $M_{BH}-M_*$ relations and the BH ``fundamental
plane''
\citep{silk1998,king2003,murray2005,di-matteo2005,sazonov2005,hopkins2007},
(2) to shut off star formation in elliptical galaxies (e.g., by
blowing gas out of the galaxy), thereby explaining how ellipticals
become ``red and dead'' (e.g., \citealt{springel2005b,ciotti10}), (3)
to heat the hot intracluster plasma (ICM) in groups and clusters,
thereby suppressing cooling and star formation in these environments
(e.g., \citealt{tabor1993,ciotti1997,croton2006}), and (4) to help
explain ``cosmic downsizing,'' namely the fact that both star
formation and AGN activity reside in progressively lower mass halos at
lower redshifts (e.g., \citealt{scannapieco2005}).

It is plausible that AGN perform the roles desired of them, but this
is by no means certain.  Understanding whether this is indeed the case
requires developing more sophisticated theoretical models that can be
compared quantitatively to observations.  There are several key
theoretical problems that must be addressed in order to better
understand the role of massive BHs in galaxy formation, and to
understand the properties of massive BHs themselves.  The first is the
problem of AGN fueling, i.e., how is gas transferred from galactic
scales ($\sim 0.1-1$ kpc) to the vicinity of the massive BH ($\lesssim
0.1$ pc)?  A second key problem is the problem of AGN feedback: how
do energy and momentum generated by accretion onto a central BH --
in the form of radiation and outflows -- couple to the surrounding
gas, and how does this affect star formation and the growth of the BH
itself?

Much of the recent work addressing the impact of BHs on galaxy
formation has used qualitatively similar physics (e.g.,
\citealt{2005MNRAS.361..776S, 2009ApJ...690..802J}).  For example,
many calculations assume that a BH of mass $M_{BH}$ will accrete mass
at a rate proportional to the Bondi rate \citep{1952MNRAS.112..195B}:  \begin{equation}
\dot{M}_{Bondi} = \frac{4 \pi f G^2 M_{BH}^2 \rho}{c_s^3}
\label{EquationBondi}
\end{equation}  \noindent where $\rho$ is the density of the surrounding gas, $c_s$ is the sound speed of that gas, and $f \sim 10-100$ is a factor taking
into account the possible multi-phase structure of the gas and that the sphere of influence of the BH is often
not resolved \citep{2009MNRAS.398...53B}.
There is, however, little justification for using equation~\ref{EquationBondi}.  The Bondi accretion rate estimate assumes that the gas surrounding the BH is spherically symmetric.  When the gas is not
spherically distributed, the rate of angular momentum transport
determines the BH accretion rate (e.g., \citealt{shlosman1990}).  It
is generally believed that the progenitors of todays $\gtrsim L^{*}$
ellipticals are gas-rich disk galaxies, the mergers of which lead to
luminous starbursts and the growth of the central massive BHs
\citep{1988ApJ...325...74S,2005ApJ...630..705H}. Most of the gas in
disk galaxies, merging galaxies, luminous starbursts
\citep{1998ApJ...507..615D,2006ApJ...640..228T}, and nearby luminous
AGN \citep{ho2008} appears to reside in a rotationally supported disk.
There is thus no reason to expect that the spherically symmetric Bondi
rate provides a good estimate of the BH accretion rate in gas rich
galaxies.  Even in the central $\sim$ parsec of own galaxy, where the
ambient gas {\em is} hot and pressure supported, the Bondi accretion
rate fails by orders of magnitude to predict the accretion rate onto
the central BH \citep{sharma07}.

There are a number of ways that an AGN can strongly influence its
surroundings (e.g., \citealt{ostriker10b}).  Relativistic jets inject
energy into intracluster plasma and may be the primary mechanism
suppressing cooling flows in galaxy clusters \citep{mcnamara2007},
even though the details of how the energy in the jet couples to the
plasma in a volume filling way are not fully understood
\citep{vernaleo2006}.  On galactic scales, a wind from an accretion
disk around the BH can drive gas out of the galaxy (e.g.,
\citealt{king2003}) as could cosmic-ray protons produced by a radio
loud AGN \citep{socrates10}.  In addition, the AGN's radiation can
strongly affect the surrounding gas, both by Compton heating/cooling
(e.g., \citealt{sazonov2005}) and by the momentum imparted as UV
radiation is absorbed by dust grains
\citep{chang1987,1988ApJ...325...74S,murray2005}.

This diversity of feedback mechanisms can be roughly separated into
two broad classes: energy and momentum injection.  We believe that
momentum injection is the dominant mode of feedback for most of the
gas in a galaxy, largely because of the very short cooling times of
dense gas.  For example, if a BH radiates at $\sim 10^{46}$ erg
$\rm{s}^{-1}$ with a typical quasar spectrum, only gas with $n
\lesssim 1 \, \rm{cm}^{-3}$ can be heated to the Compton temperature
within $\sim 100$ pc.  However, the mean gas densities in the central
$\sim 0.1-1 \, \rm{kpc}$ of luminous star forming galaxies are $\sim
10^{3-5} \rm{cm}^{-3}$
\citep{1998ApJ...507..615D,2006ApJ...640..228T}.  At these densities,
the cooling time of gas is sufficiently short that it is unable to
retain much injected energy -- be it from the AGN's radiation or from
shocks powered by AGN outflows.  Thus it is largely the momentum
imparted by AGN outflows and by the absorption and scattering of the
AGN's radiation that dominates the impact of the AGN on dense gas in
galaxies.  Since it is the dense gas that fuels star formation and the
growth of the BH itself, it is critical to understand the impact of
momentum feedback on this gas.\footnote{These conclusions do not apply
  to dilute plasma in the intracluster or intragroup medium.  The
  densities there are sufficiently low that the plasma can be efficiently
  heated by an AGN.}

In this paper, we present simulations of major mergers of spiral
galaxies using a model for the growth of BHs that includes (1) a BH
accretion rate prescription motivated by the physics of angular
momentum transport and (2) AGN feedback via momentum injection (e.g.,
radiation pressure).  Some results of this model appear in a companion
Letter \citep{2009arXiv0909.2872D}.  The remainder of this paper is
organized as follows. Section \ref{sectionMethods} presents a summary
of our methods, including a description of the model galaxies (\S
\ref{sectionICs}), the model for star formation and the interstellar
medium (\S \ref{sectionSFR}), our BH accretion and feedback model (\S
\ref{sectionBlackHoles}) and a summary of our parameter choices (\S
\ref{sectionParam}).  Section \ref{sectionResults} shows the results
of applying this model to BH growth and star formation in major
mergers of gas-rich galaxies.  In section \ref{sectionGalaxyBH} we
show that our model of BH growth and feedback produces a reasonably
tight $M_{BH}-\sigma$ correlation similar to that observed.  Finally,
in section \ref{sectionDiscussion} we discuss our results and compare
our approach to previous models in the literature.  Appendix
\ref{resolution} presents resolution tests for our fiducial simulation
while Appendix \ref{sectionAppendix} presents some of the tests used
to verify the BH accretion and feedback models that we have
implemented.

\vspace{-0.5cm}
\section{Methodology}
\label{sectionMethods}

We use a non-public update of the TreeSPH code GADGET-2
\citep{2005MNRAS.364.1105S} provided by V. Springel to perform
simulations of equal-mass mergers of galaxies.  This version of the
code includes the effective star formation model of
\cite{2003MNRAS.339..289S} but contains no AGN feedback physics.  We
modified the code further to implement models for massive BH growth
and AGN feedback.  The details of the simulations are described in the
following subsections.  The Appendices present resolution tests and
some of the tests we performed to verify our implementation of the BH
accretion and feedback model.

\vspace{-.3cm}
\subsection{Initial Conditions and Galaxy Parameters}
\label{sectionICs}

Each model galaxy used in our major merger simulations is similar to 
those in \cite{2005MNRAS.361..776S}.  They include a spherical halo of
collisionless dark matter, a centrifugally supported disk of gas and
stars, a stellar bulge, and a central point mass representing a black
  hole.  The code used to generate the initial conditions was
provided by V. Springel and is identical to that used in
\cite{2005MNRAS.361..776S} except for one change that will be
described below.

Table~1 lists the relevant galaxy and simulation parameters for the key
merger simulations we focus on in this paper.  The simulations are all
major mergers of equal mass galaxies. The fiducial simulation (top entry)
assumes a mass of $1.94\times 10^{12} M_\odot$ for each merging galaxy, of
which 4.1\% is assigned to the gas and stars in the disk, 1.36\% is
assigned to the stars in the bulge, and the rest is in a dark matter halo.
The initial mass fraction of gas in the disk is $f_g=0.1$.  This
run uses a total of $N_p=1.6\times 10^6$ particles with $6\times 10^5$ dark
matter particles, $2\times 10^5$ particles each in the gaseous and stellar
disk, and $10^5$ particles for the stellar bulge. This run has a
Plummer equivalent gravitational force softening of $\epsilon = 47$ pc.

\begin{table*}
\centering
\begin{minipage}{180mm}
\caption{Simulation Parameters}
\label{TableRunParam}
\begin{tabular}{lcccccccccccc}
  \hline
  Run Name & $M_{tot}$ & $f_{g,0}$ & $\frac{M_b}{M_d}$ & $N_p$ & $\epsilon$ & $\frac{R_{acc}}{\epsilon}$ & $\alpha$ & $\tau$ & $M_{*,new}$ & $M_{BH,f}$ & $M_{BH,p}$ & $\sigma_f$\\
  & [$M_{fid}$]${}^{a}$ & & & [$10^6$] & [$\rm{pc}$] & & & & [$10^{10} M_{\sun}$] & [$10^{8} M_{\sun}$] & [$10^{8} M_{\sun}$] & [$\rm{km} \rm{s}^{-1}$] \\ 
  \hline
  fid & 1.0 & 0.1 & 0.33 & 1.6 & 47 & 4 & 0.05 & 10 & 1.34 & 1.49 & 1.33 & 169 \\
  fidNof$^{b}$ & 1.0 & 0.1 & 0.33 & 1.6 & 47 & 4 & 0.15 & 0 & 1.36 & 18.1 & 13.5 & 170 \\
  fid3a & 1.0 & 0.1 & 0.33 & 1.6 & 47 & 4 & 0.15 & 10 & 1.34 & 1.03 & 0.90 & 168 \\
  fid6a & 1.0 & 0.1 & 0.33 & 1.6 & 47 & 4 & 0.3 & 10 & 1.35 & 0.86 & 0.77 & 167 \\
  fidTau & 1.0 & 0.1 & 0.33 & 1.6 & 47 & 4 & 0.05 & 3 & 1.36 & 5.05 & 4.31 & 163 \\
  fidt25 & 1.0 & 0.1 & 0.33 & 1.6 & 47 & 4 & 0.05 & 25 & 1.35 & 0.39 & 0.35 & 169 \\
  fid8eps & 1.0 & 0.1 & 0.33 & 1.6 & 47 & 8 & 0.05 & 10 & 1.35 & 2.70 & 1.76 & 163 \\
  fidafg & 1.0 & 0.1 & 0.33 & 1.6 & 47 & 4 & *$^{c}$ & 10 & 1.32 & 1.21 & 1.02 & 169 \\
  fidq2$^d$ & 1.0 & 0.1 & 0.33 & 1.6 & 47 & 4 & 0.05 & 10 & 1.30 & 1.40 & 1.16 & 168 \\
  fidq07$^e$ & 1.0 & 0.1 & 0.33 & 1.6 & 47 & 4 & 0.05 & 10 & 1.32 & 1.52 & 1.36 & 164 \\
  big & 3.0 & 0.1 & 0.33 & 1.6 & 68 & 4 & 0.05 & 10 & 3.08 & 6.24 & 5.27 & 232 \\
  big6a & 3.0 & 0.1 & 0.33 & 1.6 & 68 & 4 & 0.3 & 10 & 4.17 & 7.86 & 5.15 & 227 \\
  mid & 0.3 & 0.1 & 0.33 & 1.6 & 32 & 4 & 0.05 & 10 & 0.39 & 0.38 & 0.26 & 115 \\
  small & 0.1 & 0.1 & 0.33 & 1.6 & 22 & 4 & 0.05 & 10 & 0.13 & 0.24 & 0.13 & 82 \\
  small6a & 0.1 & 0.1 & 0.33 & 1.6 & 22 & 4 & 0.3 & 10 & 0.13 & 0.25 & 0.24 & 84 \\
  smallq07$^e$ & 0.1 & 0.1 & 0.33 & 1.6 & 22 & 4 & 0.05 & 10 & 0.12 & 0.06 & 0.05 & 81 \\
  fg & 1.0 & 0.3 & 0.33 & 2.4 & 47 & 4 & 0.05 & 10 & 4.41 & 7.10 & 5.53 & 159 \\
  smallfg & 0.1 & 0.3 & 0.33 & 2.4 & 22 & 4 & 0.05 & 10 & 0.36 & 0.31 & 0.23 & 98 \\
  bulge & 1.0 & 0.1 & 0.20 & 1.6 & 47 & 4 & 0.05 & 10 & 1.38 & 1.44 & 1.25 & 161 \\
  LRfid & 1.0 & 0.1 & 0.33 & 0.16 & 102 & 4 & 0.05 & 10 & 1.34 & 1.65 & 0.93 & 164 \\
  MRfid & 1.0 & 0.1 & 0.33 & 0.48 & 70 & 4 & 0.05 & 10 & 1.35 & 2.92 & 2.40 & 168 \\
  MRfidNof$^{b}$ & 1.0 & 0.1 & 0.33 & 0.48 & 70 & 4 & 0.15 & 0 & 1.34 & 13.5 & 11.4 & 167 \\
  LRfidNof$^{b}$ & 1.0 & 0.1 & 0.33 & 0.16 & 102 & 4 & 0.15 & 0 & 1.31 & 13.1 & 11.4 & 175 \\
  fidvol & 1.0 & 0.1 & 0.33 & 1.6 & 47 & 8.62 & 0.05 & 10 & 1.39 & 3.22 & 2.45 & 164 \\
  MRfidvol & 1.0 & 0.1 & 0.33 & 0.48 & 70 & 5.97 & 0.05 & 10 & 1.36 & 3.30 & 1.92 & 164 \\
  \hline
  \footnotetext[1]{$M_{fid} = 3.88 \times 10^{12} M_{\sun}$.}
  \footnotetext[2]{These runs had no AGN feedback.}
  \footnotetext[3]{$\alpha$ was set by the gas fraction within $R_{acc}$  using $\alpha = 3 f_g^2$.}
  \footnotetext[4]{The ISM equation of state is defined using $\qeos = 0.2$ (see \S \ref{sectionSFR}).}
  \footnotetext[5]{The ISM equation of state is defined using $\qeos = 0.07$ (see \S \ref{sectionSFR}).}
\end{tabular}

\medskip Columns are defined as follows: $M_{tot}$ is the total mass in the
simulation, $f_{g,0}$ is the initial gas fraction of the disk, $M_b/M_d$ is
the bulge to disk mass ratio, $N_p$ is the total number of particles used
in the simulation, $\epsilon$ is the Plummer equivalent gravitational force
softening, $R_{acc}$, $\alpha$ and $\tau$ are the parameters of the BH
accretion and feedback model (\S \ref{sectionBlackHoles}), $M_{*,new}$ is
the total mass of new stars formed during the simulation, $M_{BH,f}$ and
$M_{BH,p}$ are the masses of the BH at the end of the simulation and after
the peak of accretion (defined to be when the accretion rate drops to one
tenth its maximum value), respectively, and $\sigma_f$ is the stellar velocity
dispersion of the merger remnant (\S \ref{sectionGalaxyBH}).

\end{minipage}
\end{table*}

To test the dependence of the results of our fiducial simulation on the
model and simulation parameters, we have run a number of additional
simulations, varying the gas fraction ($f_g=0.3$ vs 0.1), bulge-to-disk
mass ratio (0.2 vs 0.33), total galaxy mass (from 0.1 to 3 of the fiducial
value), simulation particle number (from $N_p=1.6\times 10^5$ to $2.4\times
10^6$), force softening ($\epsilon=22$ to 102 pc), as well as the
parameters in the black hole model (described in \S~2.4 below).

  We use a \cite{1990ApJ...356..359H} density profile for the
  structure of the dark matter halo:
\begin{equation}
\rho_{halo}(r) = \frac{M_{halo}}{2\pi} \frac{a}{r(r+a)^3}.
\label{hernqprof}
\end{equation}
\noindent The scale length $a$ of the halo is set by requiring that
the halo enclose the same mass within the virial radius as an NFW
profile, and that the densities match at small radii.  These
conditions yield a relationship among the halo scale length, $a$,
the corresponding NFW scale length, $r_s$, and the concentration of
the NFW halo, $c$ \citep{1996ApJ...462..563N, 2005MNRAS.361..776S}:
$a = r_s \{2[\ln{(1 + c)}-c/(1+c)]\}^{1/2}$.
The halos used in this work all have a concentration of $c = 9$.

The stellar and gaseous disks both initially have exponential surface
density profiles:
\begin{equation}
\Sigma(R) = \frac{M_i}{2 \pi R_d^2} \exp \left(-\frac{R}{R_d}\right)
\label{EquationExponentialSurfDens}
\end{equation}

\noindent where $M_i$ is the total mass of the component of interest
and $R_d$ is the disk scale length, which is initially the same for
the stellar and gaseous disks.  The disk scale length for the fiducial
simulation is $R_d = 3.5$ kpc, which corresponds to the disk having
approximately the same angular momentum per unit mass as a halo with a
spin parameter of $0.033$.  For simulations with different disk
masses, we use $R_d \propto M_d^{1/3}$, which is consistent with the
observed relation \citep{2003MNRAS.343..978S}.  The stellar disk's
vertical structure is given by the standard ${\rm sech}^2(z/z_0)$
profile, where the vertical scale height $z_0$ is initially set to
$z_0 = R_d/5$ at all radii. Unlike the stellar disk, the gaseous
disk's vertical structure is determined by hydrostatic equilibrium
given the assumed sound speed/equation of state of the gas (discussed
below).  Setting up this initial vertical hydrostatic equilibrium
requires an iterative procedure that is described in
\cite{2005MNRAS.361..776S}.

The stellar bulges also have Hernquist density profiles.  The scale length
of the bulge $R_b$ is specified as a fraction of the disk scale length,
$R_d$. In the fiducial simulation, $R_b = R_d/5$.  For different bulge
masses, we use the scaling relation $R_b \propto M_b^{1/2}$, which is
motivated by the observed mass-radius relation of elliptical galaxies
\citep{2003MNRAS.343..978S}.

In our simulations, two galaxies with identical structure are placed
on a prograde orbit.  For simulations at our fiducial mass of
$1.94\times 10^{12} M_\odot$ (for each galaxy), the initial separation
of the two galaxies' centers is $142.8$ kpc.  The orbit has
approximately zero total energy, which corresponds to an initial
velocity for each galaxy of 160 km s$^{-1}$; the velocity is directed
at an angle of $28$ degrees from the line connecting the centers of
the two galaxies.  In order to break the symmetry of the problem, the
individual spin axes of the galaxies have a relative angle of about
$41$ degrees, with one galaxy of the pair having an inclination with
respect to the orbital plane of $10$ degrees. For the simulations with
different overall masses, the orbital parameters are scaled by
$M^{1/3}$, so that the time to first passage and the time to final
merger are similar to those in the fiducial run.

\vspace{-.3cm}
\subsection{Interstellar Medium Model}
\label{sectionSFR}

The version of GADGET we use includes \cite{2003MNRAS.339..289S}'s
sub-resolution model for the interstellar medium (ISM).  This model
treats the gas as a two phase medium of cold star forming clouds and a
hot ISM.  When cooling and star formation are rapid compared to the
timescale for adiabatic heating and/or cooling (which is nearly always
the case in our calculations), the sound speed of the gas is not
determined by its true temperature, but rather by an effective sound
speed that averages over the multi-phase ISM, turbulence, etc.  The
effective sound speed as a function of density can be interpolated
freely between two extremes using a parameter $\qeos$. At one extreme,
the gas has an effective sound speed of $10 \, {\rm km\,s^{-1}}$,
motivated by, e.g., the observed turbulent velocity in atomic gas in
nearby spirals; this is the ``no-feedback'' case with $\qeos=0$.  The
opposite extreme, $\qeos=1$, represents the ``maximal feedback''
sub-resolution model of \citet{2003MNRAS.339..289S}, motivated by the
multiphase ISM model of \citet{mckee77}; in this case, $100\%$ of the
energy from supernovae is assumed to stir up the ISM.  This equation
of state is substantially stiffer, with effective sound speeds as high
as $\sim200\,{\rm km\,s^{-1}}$.  Varying $\qeos$ between these two
extremes amounts to varying the effective sound speed of the ISM, with
the interpolation
\begin{equation}
c_{s} = \sqrt{\qeos\,c_{s}^{2}[q=1] + (1-\qeos)\,c_{s}^{2}[q=0]}\ .
\label{eqn:qeos}
\end{equation}

In addition to this effective equation of state, GADGET models star
formation by stochastically converting gas particles into star
particles at a rate determined by the gas density,
\begin{equation}
\dot{\rho}_{SF} = \frac{1-\beta}{t_{*}^0 \rho_{th}^{1/2}} \rho^{1/2} \rho_c \propto \rho^{3/2}
\label{equationSFRModel}
\end{equation}
\noindent where $\beta = 0.1$ is the fraction of the mass of a stellar
population returned to the ISM by stellar evolution.  The parameter
$t^0_{*}$ is the characteristic timescale for gas to be converted into
stars at the threshold density $\rho_{th} = 0.092$ cm$^{-3}$; $\rho_c
\approx \rho$ is the density of the cold clouds, which is related to
the density of the SPH particle by equations~(17) and (18) of
\cite{2003MNRAS.339..289S}.  For a given gas equation of state, the
parameters in equation~\ref{equationSFRModel} can be adjusted to
produce a global star formation law similar to the observed
Kennicutt-Schmidt relations \citep{2005MNRAS.361..776S}.

For parameters in the equation of state model that have been used in
previous work \citep{2005MNRAS.361..776S} -- $T_{SN} = 4 \times 10^8$
K, $A_0 = 4000$, $t^0_{*} = 8.4$ Gyr and $q_{EOS} = 0.5$ -- we find
that the model overpredicts the sound speed relative to the observed
``turbulent'' velocities of galaxies, i.e., the non-thermal line widths
(see Fig. 1 of \citealt{hq10} for a compilation of relevant data).
For instance, the above model parameters imply $c_s \sim 30$ km
s$^{-1}$ at $n \sim 1$ cm$^{-3}$ and $c_s \sim 110$ km s$^{-1}$ at $n
\sim 10^3$ cm$^{-3}$.  These values are too large by a factor of $\sim 2-3$
compared to the random velocities inferred from atomic and molecular
line observations \citep{1998ApJ...507..615D}.  To account for this,
we set $\qeos = 0.5$ and then modified GADGET by reducing the pressure
everywhere by a factor of $10$.  This reduces the effective sound
speed by a factor of $\sim 3$ and is thus more consistent with
observations.  This reduction in ISM pressure is also used in the
initial conditions when setting up vertical hydrostatic equilibrium
for the gas.  Changing the pressure requires changing the equation of
state parameters to $T_{SN} = 6.6 \times 10^8$ K, $A_0 = 6600$, and
$t_{*}^0 = 13.86$ Gyr to maintain an average star formation rate of $1
\, M_{\sun}$ yr$^{-1}$ for an isolated galaxy with our fiducial Milky Way
like mass. In \S \ref{sec:ISM} we compare our fiducial calculations
with this reduction in pressure to models with smaller values of
$\qeos$, $0.07$ and $0.2$; these also have smaller ``sound speeds''
more comparable to the observed random velocities of galaxies.

The reduction in the sound speed decreases the Jeans length and mass,
making it numerically more prohibitive to resolve these critical
scales.  For the simulations presented here, we are careful to use
sufficient numbers of particles so that the Jeans length and mass are
always adequately resolved.  The higher gas fraction simulations
require a higher particle number as a result (see Table
\ref{TableRunParam}).  The reduction in sound speed also makes it more
likely that the gas will fragment by gravitational instability into
clumps (ala molecular clouds), as we shall discuss in detail later.
This fragmentation is real, not numerical; artificially increasing the
sound speed to eliminate it is not necessarily physical and could give
incorrect results.  On the other hand, we do not include sufficient
physics in our ISM model to describe the formation and disruption of
molecular clouds so our treatment of the resulting clumping is also
not correct. In \S \ref{sec:ISM} we discuss which of our results are
the most sensitive to uncertainties related to local gravitational
instability in the ISM.

\vspace{-0.4cm}
\subsection{Black Hole Accretion and Feedback}
\label{sectionBlackHoles}

\subsubsection{Black Hole Accretion Model}

We include a BH as an additional collisionless particle at the center
of each galaxy. We model the accretion of the surrounding gas onto the
BH, via the transport of angular momentum, using
\begin{equation}
\dot{M}_{visc} = 3 \pi \alpha \Sigma \frac{c_s^2}{\Omega}
\label{mdotvisceqn}
\end{equation}
\noindent where $\Sigma$ is the mean gas surface density, $\Omega$ is
the rotational angular frequency, and $\alpha$ is the dimensionless
viscosity (a free parameter of our model).  We compute $\Sigma$ and
$c_s$ by taking an average of the properties of the SPH particles in a
sphere of radius $R_{acc}$ centred on the BH.  The radius $R_{acc}$ is
typically set equal to four times the gravitational force softening
length, i.e., $R_{acc} = 4 \epsilon$, although we explore alternate
choices as well.  We find that estimating the rotation rate using
$\Omega^2 \simeq GM(<R_{acc})/R_{acc}^3$ is more numerically robust
than actually calculating the rotation and angular momentum of the gas
particles within $R_{acc}$.

Although equation~(\ref{mdotvisceqn}) is reminiscent of the alpha
prescription of \cite{1973A&A....24..337S}, in our formulation
$\alpha$ characterizes not only the efficiency of angular momentum
transport, but also the uncertainty due to the fraction of the
inflowing gas that is turning into stars vs. being accreted onto the
central BH.  The physical mechanisms driving gas from $\sim$ kpc to
$\sim 0.1$ pc are not fully understood, but non-axisymmetric
gravitational torques are likely responsible
\citep{1989Natur.338...45S,hq10}.  Using numerical simulations that
focus on the nuclei of galaxies (from $\sim 0.1-100$ pc) \citet{hq10}
simulate the conditions under which there is significant gas inflow to
$\lesssim 0.1$ pc.  They argue that the net accretion rate is not a
strong function of the gas sound speed (unlike {\em both} eqns
\ref{EquationBondi} and \ref{mdotvisceqn}) because non-axisymmetric
gravitational perturbations produce orbit crossing and strong shocks
in the gas.  The resulting inflow rate depends primarily on the
non-axisymmetry in the potential, rather than the thermodynamics of
the gas.  Nonetheless, equation (\ref{mdotvisceqn}) evaluated at $\sim
100$ pc and with $\alpha \sim 0.1$ approximates the accretion rate at
small radii in their simulations, albeit with substantial scatter
(factor of $\sim 10$).  Given that one of our key results discussed in
\S \ref{sectionResults} is that the accretion rate is not sensitive to
the exact value of $\alpha$, we believe that equation
(\ref{mdotvisceqn}) is sufficient for the exploratory calculations in
this paper.

\vspace{-0.3cm}
\subsubsection{Mass of the Black Hole Particle}

In our galaxy merger simulations, the two BHs are initially far apart
but approach each other in the late stages of the merger.  When the BH
particles have a separation of less than $R_{acc}$ we consider them to
have merged.  When this occurs, we sum the individual masses of the
two BH particles and set one of the particles to have this mass.  This
particle is then moved to the center of mass of the two BH system and
given the velocity of the center of mass frame. The other BH particle
is removed from the region.

The BH particles are subject to stochastic motion due to interaction with
the stellar and gaseous particles, which leads to inaccuracy in the
position of the BH and noise in the estimate of the accretion rate.  To
reduce this numerical ``Brownian'' motion, the BH particles are given a
large ``tracer'' mass of $2 \times 10^8 M_{\sun}$ for the fiducial
simulation, and scaled with the overall mass for other simulations.  As a
result, the BH particle is a factor $\sim 100$ more massive than the halo
particles, and a factor $\sim 10^4$ more massive than the stellar and
gaseous particles.  We artificially increase the BH particle mass solely to
reduce numerical relaxation effects.  This does not result in spurious
dynamical effects on the central stars, gas, and dark matter since the BH's
sphere of gravitational influence extends to $\lesssim 10$ pc for the
fiducial simulation, which is significantly smaller than our typical force
softening of $\sim 50$ pc.

For the results presented below, the ``real'' mass of the BH ($\equiv
M_{BH}$) is computed by integrating the accretion rate
of equation~(\ref{mdotvisceqn}) in time.  The gas particles are not removed as
the BH mass increases.  Instead, the gas particles have an additional
label that tracks whether or not they have been ``consumed.''
We track how much mass the BH should have consumed via accretion at a
given time, and the mass of gas that has been consumed.  When there is
a mis-match, we tag a number of gas particles within $R_{acc}$
(chosen at random) as ``consumed'' until the total mass accreted by
the BH is correct.  Particles that have been consumed no longer
contribute to the accretion rate estimate, even if they are inside
$R_{acc}$.  This implementation prevents any gas particle from providing more than
its mass to the integrated mass of the BH.

\vspace{-0.15cm}
\subsubsection{Feedback from the Black Hole}

In our simulations, the AGN is assumed to couple to the surrounding
gas by depositing momentum into the gas, directed radially away from
the BH.  This crudely approximates the effects of (1) strong outflows
and/or cosmic-ray pressure produced by the AGN
\citep{king2003,socrates10} and (2) radiation pressure produced by
the absorption and scattering of the AGN's radiation by dust in the
ISM \citep{murray2005}. We focus on the latter when motivating the
parameters used in our models.  

To accurately account for the impact of the AGN's radiation on gas in
its host galaxy would require a radiative transport calculation, which
is beyond the scope of the current work.  Instead, we model this
radiation pressure by depositing a total momentum per unit time of
\begin{equation}
  \dot{p} = \tau \frac{L}{c} \quad \mbox{ where } L = {min}\left(\eta \dot{M}_{visc} c^2, L_{Edd}\right)
\label{momdepeqn}
\end{equation}
\noindent radially away from the BH into the SPH particles within a
distance of $R_{acc}$ of the BH particle. This momentum is equally
distributed among the particles so that each particle experiences the same
acceleration.  We use a radiative efficiency of $\eta = 0.1$ in all
simulations.  The physical picture behind our feedback model in equation
(\ref{momdepeqn}) is that the feedback is produced by the absorption of the
ultraviolet light from the AGN by dust in the surrounding gas, and the
subsequent reemission of infrared radiation that must diffuse its way out
of the nuclear region.  As described shortly, the parameter $\tau$ is the
total infrared optical depth of the nuclear region.

To motivate equation~(\ref{momdepeqn}) in more detail, we note that AGN
radiate most of their radiation in the ultraviolet.  The opacity of
dusty gas to UV radiation is $\kappa_{UV} \sim 10^3$ cm$^2$ g$^{-1}$,
so that only a surface density of $\sim 10^{-3}$ g cm$^{-2}$ is
required to absorb the UV radiation.  This is far less than the
typical radial column density of gas in the central $\sim 0.1-1$ kpc
of luminous star forming galaxies, galaxy mergers, or our simulations
(see Fig. ~\ref{FigureSigmaFiducial} below).  As a result, the UV
radiation is efficiently absorbed, except perhaps along polar lines of
sight.  The absorption and scattering of the UV radiation deposits a
momentum per unit time of $L/c$ into the ambient gas, assuming for
simplicity that all of the UV radiation is absorbed.  If the infrared
optical depth is $\gtrsim 1$, the infrared radiation re-emitted by the
dusty gas must diffuse out through the nuclear region; doing so
deposits an additional momentum per unit time of $\tau L/c$, where
$\tau \sim \kappa_{IR} \Sigma$ is the infrared optical depth and
$\kappa_{IR} \sim$ few-10 cm$^2$ g$^{-1}$ is the infrared opacity for
the radiation temperatures of interest $\sim 100-1000$ K.  The net
force due to the UV and infrared radiation is thus 
$ \dot p \sim  (1 + \tau ) L/c \simeq \tau L/c$,
i.e. equation~(\ref{momdepeqn}),
for $\tau \gtrsim 1$, which is valid in our calculations near the
peak of activity when the BH gains most of its mass.

In our calculations we use a constant value of $\tau$ rather than a
time variable $\tau$ given by $\tau = \kappa_{IR} \Sigma$. Given the
simplicity of our feedback model relative to a true radiative transfer
calculation, this is not an unreasonable approximation.  It is also
easier to isolate the effects of varying $\tau$ when it is constant in
time.

As noted above, we apply the force in equation (\ref{momdepeqn}) to
all particles within a distance $R_{acc}$ of the BH. A more accurate
treatment would be to apply the force out to the point where the
column is $\sim \kappa_{IR}^{-1}$, i.e., to where the optical depth to
infinity is $\sim 1$.  At many times, however, this radius is
unresolved.  Moreover, it is possible that the photons diffuse
primarily along the rotation axis of the gas, rather than in the
orbital plane.  As a result, the radiation pressure force will be
applied primarily at small radii.  This is why we apply the force only
within $R_{acc}$.  One consequence of this is that the number of SPH
particles experiencing the feedback, $N$, will change as gas moves in
and out of $R_{acc}$.  Thus, the strength of feedback felt by an
individual particle will change with time.  However, because the SPH
particles are collisional, they readily share this momentum with
neighboring gas particles. In test problems described in Appendix B
the effects of our feedback model are essentially independent of $N$
and $R_{acc}$.  The results are not quite so clean in our full
simulations (see \S \ref{sec:BHmodel} and Appendix \ref{resolution}),
but nonetheless none of our major results depend sensitively on the
region over which the feedback force is applied.

One might worry that if the number of particles within $R_{acc}$ were too
small, the momentum supplied to a single particle would become large enough
to artificially accelerate the particle to the escape velocity.  The
minimum $N$ required to avoid this is actually quite modest for the range
of luminosities in our calculations, and for the simulations presented here
this concern is never an issue (although it is for some of the test
problems in Appendix B).

\vspace{-0.4cm} 
\subsection{Parameter Choices for the Black Hole Model}
\label{sectionParam}
                                                                                               
Our model for BH growth and feedback contains three free parameters:
(1) $\alpha$ determines the magnitude of the accretion rate onto the
BH; (2) $\tau$ determines the total radiation pressure force produced
by accretion onto the BH; it is roughly the optical depth to the far
IR in the nuclear region; and (3) $R_{acc}$ is the radius of the
spherical region within which the accretion rate is determined and the
feedback is applied.  Our fiducial values for these parameters are
$\alpha = 0.05$, $\tau = 10$, and $R_{acc} = 4 \, \epsilon$ (where
$\epsilon$ is the gravitational force softening).  We now motivate
these particular choices.

The fiducial value of the viscosity used in this work is $\alpha =
0.05$, motivated by the rough consistency between the resulting $\dot
M$ and \citet{hq10}'s numerical simulations of gas inflow from $\sim
100$ pc to $\sim 0.1$ pc (although there is factor of $\sim 10$
scatter in the latter that is not captured here).  \citet{hq10}'s
calculations in fact require a more complicated subgrid accretion
model that depends on additional parameters such as the bulge to disk
ratio of the galaxy (because this influences the strength of
non-axisymmetric torques); this will be explored in more detail in
future work.  In addition to $\alpha = 0.05$, we also carried out
simulations with $\alpha = 0.15$ and $\alpha = 0.3$, and found no
significant differences, for reasons explained below.

We use a constant value (with time) of $\tau = 10$ in most of our
simulations.  This is motivated by far infrared opacities of
$\kappa_{IR} \sim 3-10$ cm$^{2}$ g$^{-1}$ and surface densities of
$\Sigma \sim 1-10$ g cm$^{-2}$ within $R_{acc}$ during the peak of
activity in our simulations.  These surface densities are also
consistent with those directly measured in the nuclei of
ultra-luminous infrared galaxies \citep{1998ApJ...507..615D}.  Given
the uncertainties associated with the radiative transfer of far
infrared photons in galactic nuclei, it is not possible to more
accurately estimate the effective value of $\tau$ without detailed
radiative transfer calculations.  As we shall demonstrate explicitly,
however, the exact value of $\tau$ is also not that critical for the
qualitative effects of AGN feedback; the value of $\tau$ does,
however, strongly affect the final value of the BH mass.

In choosing a value for $R_{acc}$, we must satisfy $R_{acc} > \epsilon$ in
order to avoid numerical artifacts.  In addition, we find that the BH
particle remains within $4 \epsilon$ of the centre of mass of the system at
nearly all times, but it can wander around within this region.  As a
result, $4 \epsilon$ is the smallest we can make $R_{acc}$ without having
noise induced by the BHs motion.  This choice corresponds to several
hundred pc in our typical simulation.  Larger values of $R_{acc}$ are
unphysical because (1) the accretion rate should only depend on the gas
close to the BH; i.e., the transport of gas from, for example, $\sim 8
\epsilon$ to $\sim 2 \epsilon$ is presumably adequately described by our
simulations so we should not try to also account for this in our subgrid
model, and (2) the radiation pressure force produced by the AGN (and the
re-radiated infrared photons) is likely concentrated at relatively small
radii, for the reasons described in \S \ref{sectionBlackHoles}.

\vspace{-0.6cm}
\section{Galaxy Merger Simulations}
\label{sectionResults}

Table~\ref{TableRunParam} summarizes the simulations we focus on in
this paper, including the resolution, the parameters that specify the
initial conditions for the merging galaxies, the parameters that
specify the BH accretion and feedback models, and the final properties
of the merger remnants (stellar and BH mass and velocity dispersion).
We begin by describing the results from our fiducial simulation (top
row in Table~\ref{TableRunParam}) and then discuss simulations that
vary a single parameter of the feedback model relative to the fiducial
run.  We have also performed simulations at different overall galactic
mass scales, initial gas fractions, and numerical resolution.  The
latter resolution tests are presented in Appendix A.

\vspace{-0.35cm}
\subsection{The Fiducial Simulation}
\label{sec:fid}

\begin{figure}
\includegraphics[width=84mm]{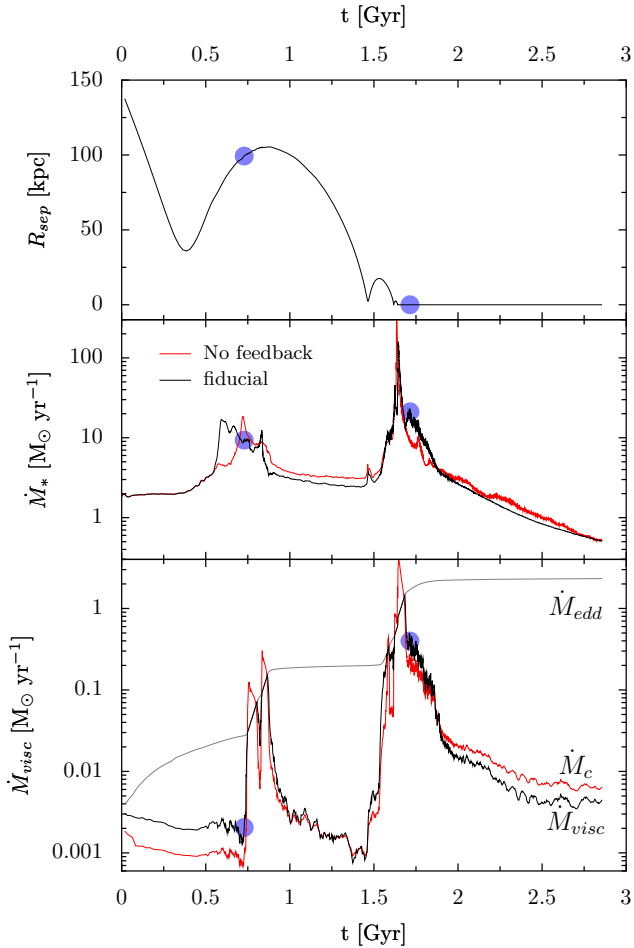}
\caption{ \emph{Top:} The separation of the black hole particles as a
  function of time in the fiducial simulation.  The blue circles label
  the times of the images shown in Figure~\ref{figureSFRClump}.
  \emph{Middle:} The star formation rate as a function of time for the
  fiducial simulation (black) and for the run with no feedback (red;
  run fidNof).  \emph{Bottom:} The viscous accretion rate,
  $\dot{M}_{visc}$ (black), and Eddington rate (grey), as functions of
  time for the fiducial simulation.  The critical $\dot{M}_c$ at which
  radiation pressure balances gravity (eq.~\ref{EquationMCrit}) is
  shown within a radius of $R_{acc}$ (red; solid).  The increase in
  star formation and BH accretion after first passage ($t \sim 0.75$
  Gyr) is due to the fragmentation and inspiral of large
  gaseous/stellar clumps (Fig. \ref{figureSFRClump}), while the much
  larger increase at final coalescence is due to inflow of diffuse gas
  caused by non-axisymmetric torques. The latter physics dominates the
  total stellar and BH mass formed during the merger.}
\label{FigureFiducialMegaPlot}
\end{figure}

The top panel of Fig.~ \ref{FigureFiducialMegaPlot} shows the
separation of the BH particles for the fiducial simulation, while the
middle panel shows the total star formation rate (in both galaxies)
for simulations with (black) and without (red) BH feedback. The first
close passage of the two galaxies is around $t = 0.33$ Gyr and the
system then undergoes a few short oscillations as the BHs finally
settle into a merged state around $t = 1.65$ Gyr.  The star formation
rate increases following the first passage, with a much larger
increase in the star formation rate during the final merger of the
galaxies.  The bottom panel of Fig.~ \ref{FigureFiducialMegaPlot}
shows the BH accretion rate determined from equation~\ref{mdotvisceqn}
(black) and the Eddington accretion rate (grey; $\dot M_{edd} \equiv
L_{edd}/0.1c^2$); the initial BH mass is $1.4 \times 10^5 M_\odot$ but
as long as it is not too large $\gtrsim 10^8 M_\odot$, the precise
initial BH mass is unimportant for our conclusions.  In this and
similar plots throughout the paper, the value of $\dot{M}$ plotted
before the BHs merge is for the BH in the galaxy with the smaller
initial inclination relative to the orbital plane; the BH accretion
rate for the other galaxy is comparable to that shown here.  The
evolution of the accretion rate is similar in many of the simulations
we have carried out, with an initial period of activity after the
first passage of the merging galaxies, and another period of even
higher $\dot M$ after the final coalescence of the galaxies and BHs.
The latter active episode is when the merged BH gains most of its
mass.  In particular, the BH reaches the Eddington limit, allowing the
mass of the BH to grow exponentially for a few hundred Myr.

\cite{2009arXiv0909.2872D} showed that the BH accretion and feedback
model presented in this work leads to self-regulated BH growth, due to
a competition between the (inward) gravitational force produced by the
galaxy as a whole and the (outward) radiation pressure force produced
by the central AGN (eq. \ref{momdepeqn}) \citep{murray2005}.  For a
spherically symmetric system, equating these two forces leads to $\tau
L / c = 4 f_g \sigma^4 / G$, where $\sigma^2 = G M_t / 2 R_{acc}$,
$M_t$ is the total mass inside $R_{acc}$, and we have evaluated these
expressions within $R_{acc}$, where our accretion rate is determined
and feedback is implemented.  Equivalently, there is a critical
accretion rate $\dot{M}_{c}$, analogous to the Eddington rate, at
which the two forces balance:
\begin{equation}
\dot{M}_c =  \frac{4 f_g}{\eta \tau G c} \sigma^4.
\label{EquationMCrit}
\end{equation}

\noindent The bottom panel of Fig.~\ref{FigureFiducialMegaPlot} shows
$\dot{M}_c$ for our fiducial simulation, evaluated within $R_{acc}$ of
the BH (solid red).  Comparing $\dot M_c$ to the BH accretion rate
$\dot M_{visc}$ demonstrates that during the peak episodes of
accretion $\dot{M}_{visc} \sim \dot{M}_c$, so that radiation pressure
becomes dynamically important. Although it is certainly possible to
have accretion rates smaller than $\dot{M}_c$ when there is
insufficient gas to fuel the AGN, the accretion rate is limited to a
maximum value of $\sim \dot{M}_c$.

\begin{figure}
\includegraphics[width=84mm]{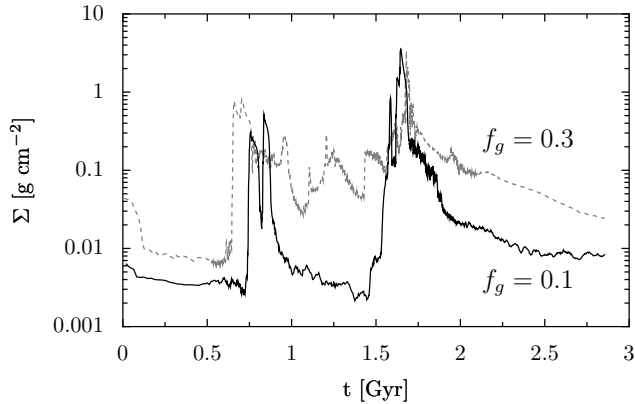}
\caption{The mean gas surface density $\Sigma$ interior to the
  accretion radius $R_{acc} = 4\epsilon = 0.19$ kpc for the fiducial
  simulation with initial gas fraction $f_g = 0.1$ (solid) and for the
  simulation with $f_g = 0.3$ (dashed; run fg).}
\label{FigureSigmaFiducial}
\end{figure}

Fig.~\ref{FigureSigmaFiducial} shows the surface density of gas
within $R_{acc} = 4 \, \epsilon = 0.19$ kpc for the fiducial
simulation and for a higher gas fraction simulation with $f_g = 0.3$.
As implied by Fig.~\ref{FigureFiducialMegaPlot}, there are two main
epochs during which significant gas is driven into the nuclei of the
galaxies: after first passage and at final coalescence.  The physical
origin of these high nuclear gas densities are, however, somewhat
different. 

\begin{figure*}
\includegraphics[width=170mm]{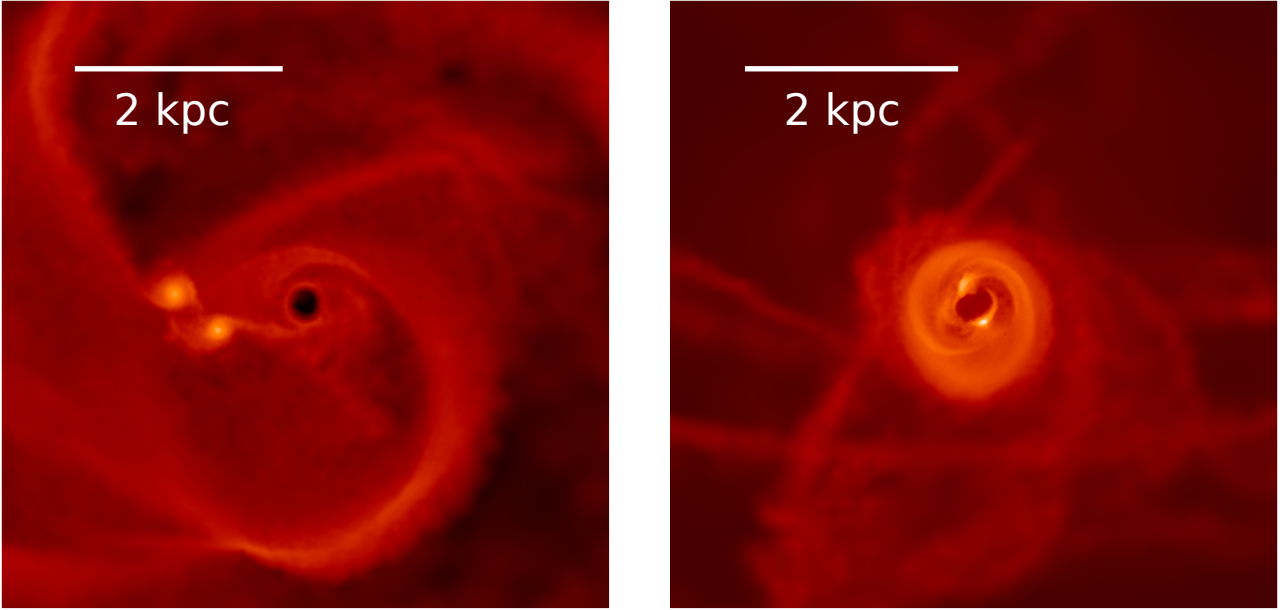}
\caption{Gas density in the vicinity of the BH for the fiducial
  simulation at $t = 0.73$ Gyr ({\em left panel}), just prior to the
  onset of significant BH accretion after the first close passage of
  the two galaxies, and $t = 1.71$ Gyr ({\em right panel}), the peak
  of star formation and BH accretion after the galaxies and BHs have
  coalesced.  The times of these images are labeled with blue circles
  in Figure~\ref{FigureFiducialMegaPlot}. In the left panel, the image
  is for the less inclined galaxy and the companion galaxy is well
  outside the image.  The images are $5.7$ kpc on a side and brighter
  color indicates a higher density.  The dark region in the center of
  each image is within $R_{acc}$ of the BH and is evacuated by BH
  feedback.  In the image just after first passage ({\em left panel}),
  the two bright white regions are gaseous/stellar clumps that
  fragmented by Toomre instability during first passage and then
  spiraled into the nucleus, fueling star formation and BH accretion.
  At final coalescence ({\em right panel}), the nuclear gas densities
  are significantly higher (see also Fig.~\ref{FigureSigmaFiducial})
  and most of the gas resides in a $\sim 1$ kpc diameter disk driven
  into the nucleus by non-axisymmetric stellar torques during the
  merger.  These images were made using SPLASH
  \citep{2007PASA...24..159P}.}
\label{figureSFRClump}
\end{figure*}

\citet{1996ApJ...464..641M} showed that the presence of a bulge like
that in our simulation suppresses a nuclear starburst after first
passage during galaxy mergers, because the bulge inhibits the
non-axisymmetric modes that drive inflow. In our fiducial simulation,
the majority of the increase in star formation after first passage is
due to gravitational instability and fragmentation of the gas, which
produces dense regions of rapid star formation.
Fig.~\ref{figureSFRClump} (left panel) shows the gas density in the
vicinity of one of the incoming black holes at $t = 0.74$ Gyr, midway
through the first peak in star formation; the companion galaxy is well
outside of this image.  Two knots of dense gas are clearly seen, both
of which will soon enter $R_{acc}$, the BH accretion and feedback
region.  These two clumps are not the only ones that form after first
passage, but they are the only clumps that survive to enter the
central region surrounding the BH.\footnote{In the simulation with a
  higher initial gas density ($f_g = 0.3$), so many fragments form at
  large radii and spiral into $R_{acc}$ that the surface density in
  the central region remains elevated from first passage until the
  merger completes at $t \sim 1.8$ Gyr (see
  Fig.~\ref{FigureSigmaFiducial}).}  Fig.~\ref{figureSFRClump} (right
panel) also shows an image of the gas density in the nuclear region at
$t = 1.71$ Gyr, near the peak of star formation and BH accretion and
after the galaxies and BHs have coalesced. At this time, the gas
density in the nuclear region is significantly higher than at first
passage (see also Fig.~\ref{FigureSigmaFiducial}) and most of the gas
resides in a $\sim 1$ kpc diameter disk.  This nuclear gas
concentration is the diffuse ISM driven in from larger radii by
non-axisymmetric stellar torques during the merger (e.g.,
\citealt{1996ApJ...464..641M}).

The galaxies in our fiducial simulation are stable when
evolved in isolation.  The merger itself drives the gas to fragment by
locally exceeding the Jeans/Toomre mass. In reality, the gas in such
clumps might disperse after $\sim$ a Myr because of stellar feedback
not included in our calculations \citep{murray10}.  This would
probably not significantly change our estimate of the star formation
rate since we are already normalized to the observed Kennicutt
relation; however, such dispersal would lead to little inflow of gas
associated with the inspiral of stellar clusters and thus would
suppress the first peak in BH accretion (see \citealt{hq10} for a more
detailed discussion).  In \S \ref{sec:ISM} we will return to these
issues and show that the total stellar mass and BH mass formed during
the merger are relatively insensitive to the details of our assumed
ISM model.

\begin{figure}
\includegraphics[width=84mm]{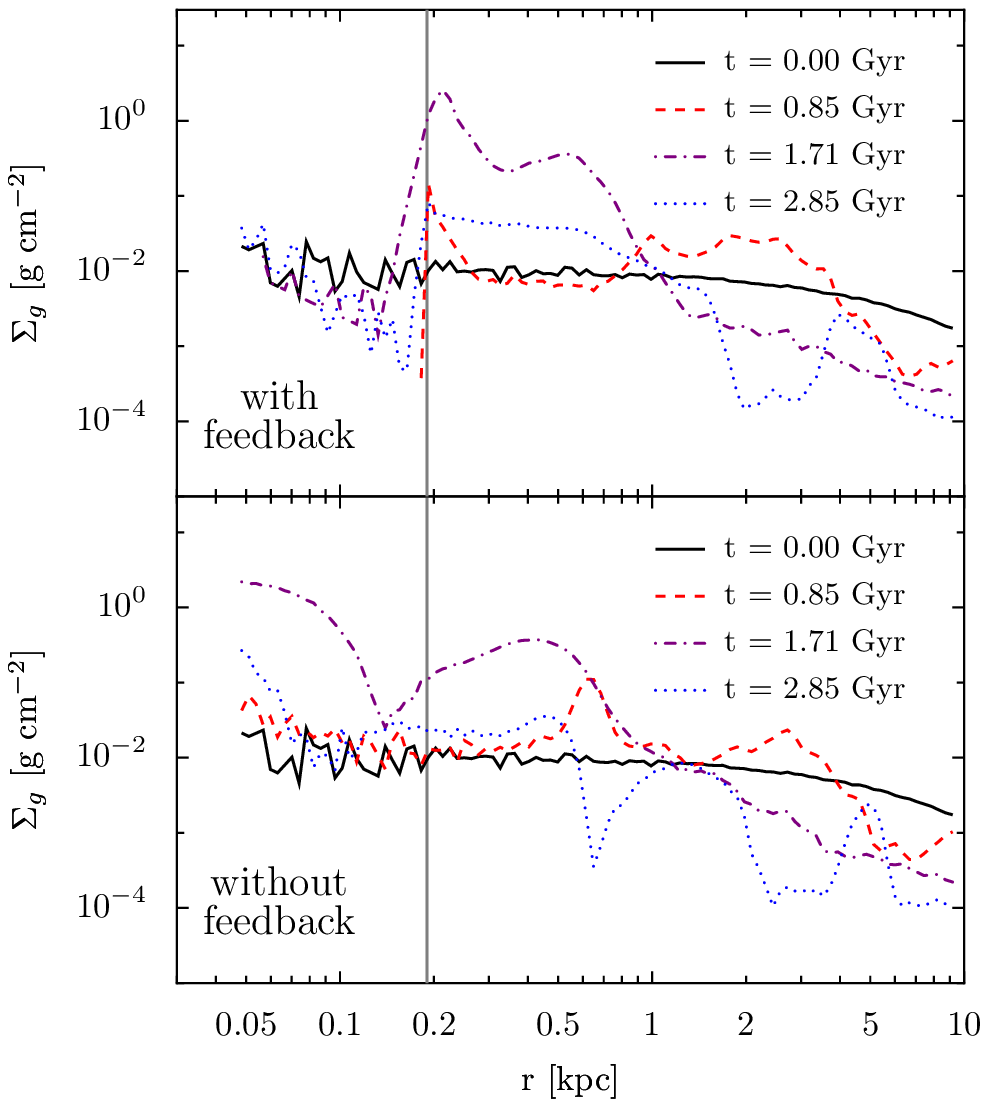}
\caption{Comparison of gas surface density ($\equiv M_g[<r]/\pi r^2$)
  versus distance from the BH in the fiducial simulation with feedback
  (top) and without feedback (bottom).  Four times are shown: $t = 0$,
  0.85 Gyr (first passage), 1.71 Gyr (peak accretion), and 2.85 Gyr
  (end of simulation). Note that the gas tends to pile up at $R_{acc}
  = 0.190$ kpc (shown by the vertical line) in the top panel.}
\label{FigureSigmaofRFiducial}
\end{figure}

Fig.~\ref{FigureSigmaofRFiducial} shows the surface density of gas
in the fiducial simulation (top panel) and for the run without
feedback (bottom panel) as a function of distance from the BH at four
times: the initial condition ($t = 0$), shortly after the first close
passage of the two galaxies ($t = 0.85$ Gyr), near the peak of
accretion ($t = 1.71$ Gyr) and at the end of the simulation ($t =
2.85$ Gyr).  Once $\dot M \sim \dot M_c$ at first passage $\sim 0.85$
Gyr, gas is driven out of the nuclear region by the AGN's radiation
pressure.  Since at the same time gravitational torques continue to
drive gas inwards, the gas begins to pile up at $\sim R_{acc}$.  The
particular radius at which the pile up occurs of course depends on our
choice of $R_{acc}$, and so the particular size of the evacuated
region should not be taken too seriously. Qualitatively, however, the
behavior in Fig. \ref{FigureSigmaofRFiducial} is reasonable: the AGN
pushes on the gas in its neighborhood until it deprives itself of
fuel.

Near the peak of activity at $t = 1.71$ Gyr, the gas surface density
in the central $R_{acc} \simeq 0.19$ kpc is a factor of $\sim 10-30$
larger in the simulations without feedback (bottom panel of
Fig. \ref{FigureSigmaofRFiducial}).  However, the gas density at large
radii $\sim 0.5$ kpc is not that different.  The radiation pressure
force from the BH thus largely affects gas in its immediate
environment, rather than the entire gas reservoir of the galaxy.
Another indication of this is that the star formation rate is very
similar in the simulations with and without feedback (middle panel of
Fig. \ref{FigureFiducialMegaPlot}).

\vspace{-0.2cm}
\subsection{Dependence on Parameters of the BH Model}
\label{sec:BHmodel}

\begin{figure*}
\includegraphics[width=180mm]{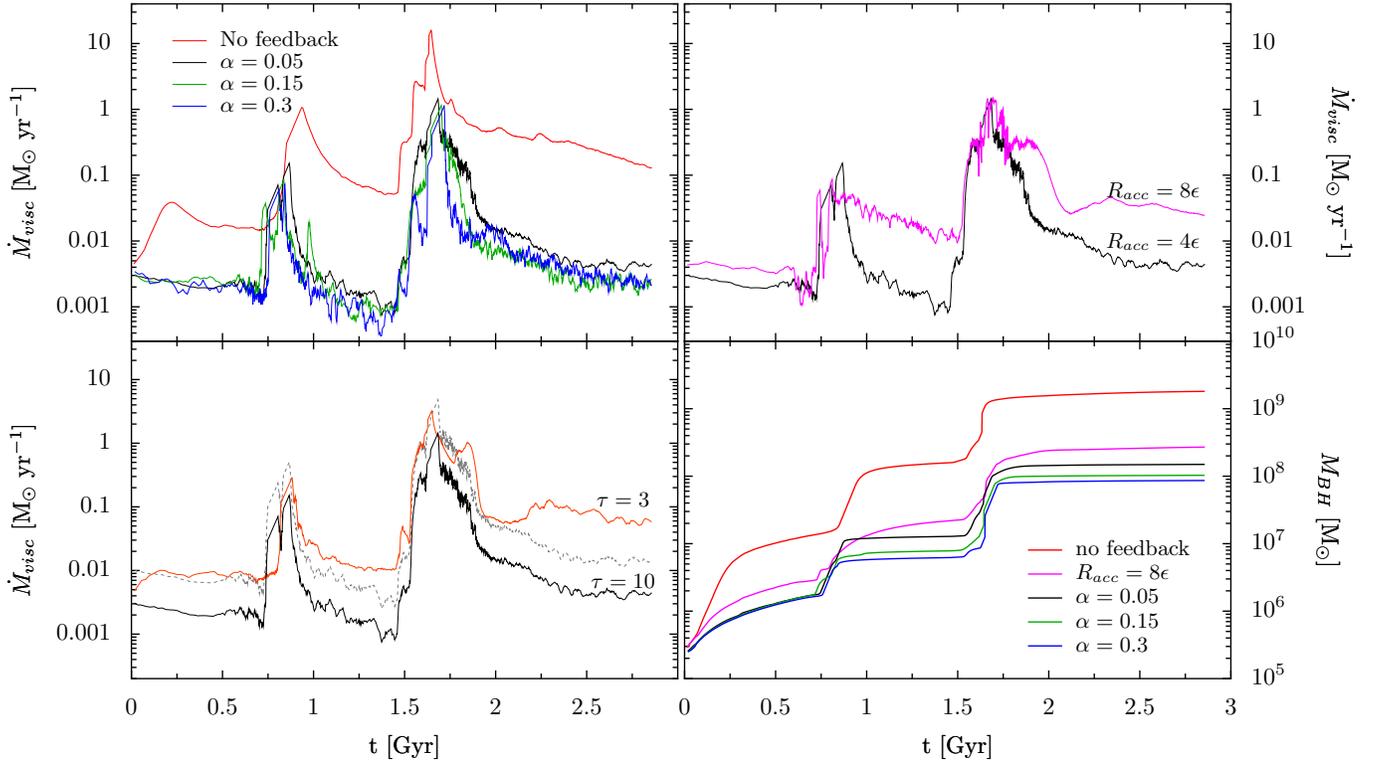}
\caption{\emph{Top Left:} Comparison of the accretion rates for the
  run without feedback (red; fidNof, $\alpha = 0.15$), and three
  runs with feedback: the fiducial simulation with $\alpha = 0.05$
  (black), the run with $\alpha = 0.15$ (green; fid3a) and the run with
  $\alpha = 0.3$ (blue; fid6a).  \emph{Top Right:} The accretion rate
  for the fiducial run (black) and the run with $R_{acc} = 8 \epsilon$
  (magenta; fid8eps).  \emph{Bottom Left:} The accretion rate for the
  fiducial run (black) and the run with $\tau = 3$ (orange; fidTau).
  Also shown is $\dot M_{visc}$ for the fiducial run increased by a
  factor of $10/3$ (dashed line), as expected from
  eq.~(\ref{EquationMCrit}).  \emph{Bottom Right:} The integrated
  black hole masses for all the runs in this Figure with $\tau = 10$.}
\label{FigureFourPanel}
\end{figure*}

The models for BH accretion and feedback used here contain uncertain
parameters.  We have defined the three relevant parameters $\alpha$,
$\tau$, and $R_{acc}$ in \S \ref{sectionParam} and motivated our fiducial
values, but it is important to explore how our results change with
variations about our fiducial parameters.

The value of $\alpha$ parameterizes the efficiency with which gas accretes
from $\sim R_{acc} \sim 190$ pc to smaller radii, encapsulating both the
efficiency of angular momentum transport and the effects of star formation
on unresolved scales.  Naively, a higher value of $\alpha$ would lead to a
more massive BH.  This is, however, not the case, because during the epochs
when the BH gains most of its mass, the accretion rate is set by the
efficiency of feedback (eq. \ref{EquationMCrit}) not by the available mass
supply (see Figs \ref{FigureFiducialMegaPlot} \&
\ref{FigureSigmaofRFiducial}).  To demonstrate this more explicitly, the
top left panel of Fig.~\ref{FigureFourPanel} compares the BH accretion
rates for three simulations with feedback, but differing values of $\alpha$
(0.05, 0.15, and 0.3), to the simulation with no feedback, which has
$\alpha = 0.15$.  The accretion histories for the three values of $\alpha$
are nearly identical.  By contrast, the accretion rate is in general much
larger in simulations that neglect feedback (and is $\propto \alpha$).
In addition to the constant $\alpha$ runs, we tested a model in which
$\alpha$ was time variable, set by the local gas fraction near the BH
(fidafg2 in Table~\ref{TableRunParam}): $\alpha = 3 f_g^2$, with $f_g$
determined within $R_{acc}$ (in practice $\alpha$ varied from $\sim 2
\times 10^{-4}-0.3$).  Although this precise functional form is
somewhat arbitrary, such a variation is motivated by analytic
arguments and numerical simulations which show that instabilities due
to self-gravity dominate the transport of gas from $\sim 100$ pc
inward \citep{shlosman1990,hq10}.  For our $\alpha = 3 f_g^2$
simulation, we find that the peak accretion rates and final BH mass
are very similar to the constant $\alpha$ simulations.  This is
consistent with our conclusion that in the limit of large fuel supply,
feedback, rather than the efficiency of angular momentum transport,
sets the rate at which the BH grows.

The parameter $\tau$ describes the efficacy of the feedback for a
given AGN luminosity.  The bottom left panel of
Fig. \ref{FigureFourPanel} compares the BH accretion rate for the
fiducial run with $\tau = 10$ (black) and a simulation with a smaller
value of $\tau = 3$ (orange).  To the extent that the accretion rate is
feedback limited and set by $\dot M_c$ in
equation~\ref{EquationMCrit}, $\dot M$ should decrease with increasing
$\tau$.  Physically, this is because larger $\tau$ leads to a larger
feedback force, which then requires a smaller accretion rate to
provide the luminosity necessary to drive away the surrounding gas.
This expectation is borne out by the simulations.  To compare the
numerical results with the scaling in equation~\ref{EquationMCrit},
the bottom left panel of Fig. \ref{FigureFourPanel} also shows $\dot
M$ for the fiducial simulation scaled by a factor of $10/3$ (dashed
line).  This scaled $\dot M$ of the fiducial simulation is in
reasonably good agreement with the $\tau = 3$ simulation, particularly
at the first and second peaks in $\dot M$, when most of the BHs mass
is accumulated.  This demonstrates that the value of $\tau$ does not
significantly affect any of the qualitative behavior of how the BH
grows, although it does determine the overall value of the BH mass.

In the majority of the simulations presented here, the size of the
region over which we apply the feedback and average the gas properties
to calculate $\dot M$, $R_{acc}$, is set to $4 \, \epsilon$.  The
rationale for this choice was given in \S \ref{sectionParam}, but it is
important to consider the effects of changing this value.  The top
right panel of Fig. \ref{FigureFourPanel} shows the mass accretion
rate for the fiducial simulation and a simulation with $R_{acc} = 8
\epsilon = 380$ pc.  The peak values of $\dot M$ and the time of the
first and second peaks are reasonably similar in the two cases.  The
principle difference is that in the simulations with the larger value
of $R_{acc}$, the feedback is less effective at clearing gas out of
the nuclear region (because the force is distributed over a larger
number of particles); this allows a higher level of $\dot M$ to be
maintained after the first passage and final coalescence.  We suspect
that the fiducial simulation better approximates what a higher
resolution calculation with radiative transfer would find, but this
remains to be demonstrated.

\begin{figure}
\includegraphics[width=84mm]{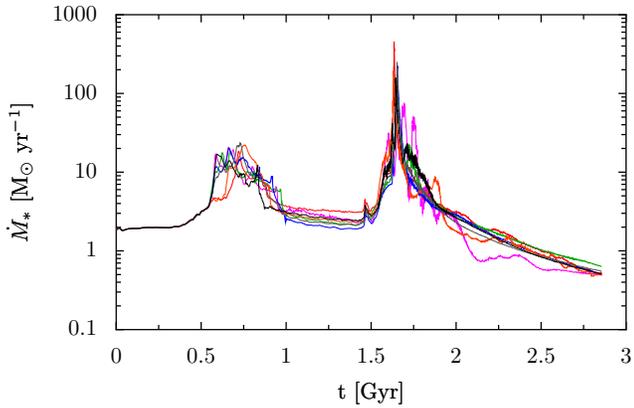}
\caption{The star formation rate for the run with no feedback (red)
  and for runs with various values of the BH accretion and feedback
  parameters: $\alpha = 0.05, 0.15, 0.3$ (black, green, blue), $\alpha =
  3 f_g^2$ (grey), $\tau = 3$ (orange), and $R_{acc} = 8 \epsilon$
  (magenta).  All of these models have very similar star formation
  histories.}
\label{FigureSFRAlphas}
\end{figure}  

The bottom right panel of Fig. \ref{FigureFourPanel} shows the integrated
BH mass as a function of time for the fiducial simulation and for the
variations in the feedback/accretion model considered in this subsection
that have the same value of $\tau$ (but different values of $\alpha$ and/or
$R_{acc}$).  The key result is that in the presence of feedback (all but
the top curve), there is only a factor of $\simeq 3$ change in the BH mass
due to differences in how we treat BH accretion and feedback.  A factor of
$6$ change in $\alpha$ leads to only a $42 \%$ change in the final BH
mass. This is because most of the BH mass is gained during the final
coalescence of the two galaxies, at which point the BH accretion
self-regulates and reaches the Eddington-like value in
equation~(\ref{EquationMCrit}).  The run without feedback (top curve), by
contrast, has a factor of $\sim 10$ larger BH mass and the BH mass would
scale linearly with the assumed value of $\alpha$.

The star formation rates for the simulations with different BH feedback
parameters are all shown in Fig.~\ref{FigureSFRAlphas} (this includes the
fiducial simulation with and without feedback and the runs with $\alpha =
0.15, 0.3, 3 f_g^2$, $\tau = 3$, and $R_{acc} = 8\epsilon$). This figure
demonstrates that the precise parameters of the BH feedback model have
little effect on the galaxy-wide properties such as the star formation
rate: the total mass of stars formed in simulations with different BH
feedback parameters differ by less than $5\%$.

In previous simulations of BH growth and feedback, AGN feedback acting
on dense gas in galaxies has been invoked to quench star formation
\citep{springel2005b}.  Our results demonstrate, however, that this is
by no means guaranteed (we refer here to `quasar' feedback on cold
dense gas, not the effect of AGN on hot dilute gas in galaxy groups
and clusters). In our calculations BH growth is self-regulated and
closely connected to the properties of the surrounding galaxy (e.g.,
eq.~\ref{EquationMCrit}).  However, the BHs dynamical influence is
centered in the galactic nucleus ($\lesssim 300$ pc); as a result, the
BH does not significantly alter the star formation history during a
merger.  In this scenario, the merger remnant can nonetheless be
relatively quiescent (``red and dead'') because the burst of star
formation uses up much of the available gas.

\vspace{-0.4cm}
\subsection{Effects of the ISM Model}
\label{sec:ISM}

Motivated by observations (e.g., \citealt{1998ApJ...507..615D}), we have
reduced the effective sound speed in GADGET's subgrid ISM model (see \S
\ref{sectionSFR}).  There is nonetheless considerable uncertainty in the
accuracy of this (or any other) subgrid model.  To study in more detail the
effects of the ISM model on our results, we performed two additional
simulations at our fiducial galaxy mass with the subgrid interpolation
parameter $\qeos = 0.2$ and $0.07$ (see eq. \ref{eqn:qeos}), and without
the factor of 10 reduction in pressure used in our fiducial simulation (an
additional simulation with $\qeos = 0.07$ at a lower galaxy mass will be
discussed in \S \ref{sectionGalaxyBH}).\footnote{We used $T_{SN} = 4 \times
  10^8$ K, $A_0 = 4000$ and $t_*^0 = 8.4$ Gyr for these calculations; these
  values are different from those in our fiducial simulation, and are
  chosen to fix the total star formation rate for our isolated fiducial
  galaxy at $1 M_{\sun}$ yr$^{-1}$.}  The three different ISM models have
$c_s$ and $Q$ within a factor of $\sim 2$ of one another at all radii, with
the $\qeos = 0.2$ model having the largest values of $c_s$ and Q, and our
fiducial model having the smallest values. The parameter $Q$ is initially
$\sim 3$ for our fiducial simulation at the disk scale length $R_d$, which
is why the merger can induce significant fragmentation of the gas
(Fig.~\ref{figureSFRClump}).  Given the limited physics included in the
subgrid model, we do not believe that it is feasible to unambiguously
conclude which of these ISM models is more realistic.  These models thus
provide an indication of the systematic uncertainty introduced by our
treatment of the ISM.

\begin{figure}
\includegraphics[width=84mm]{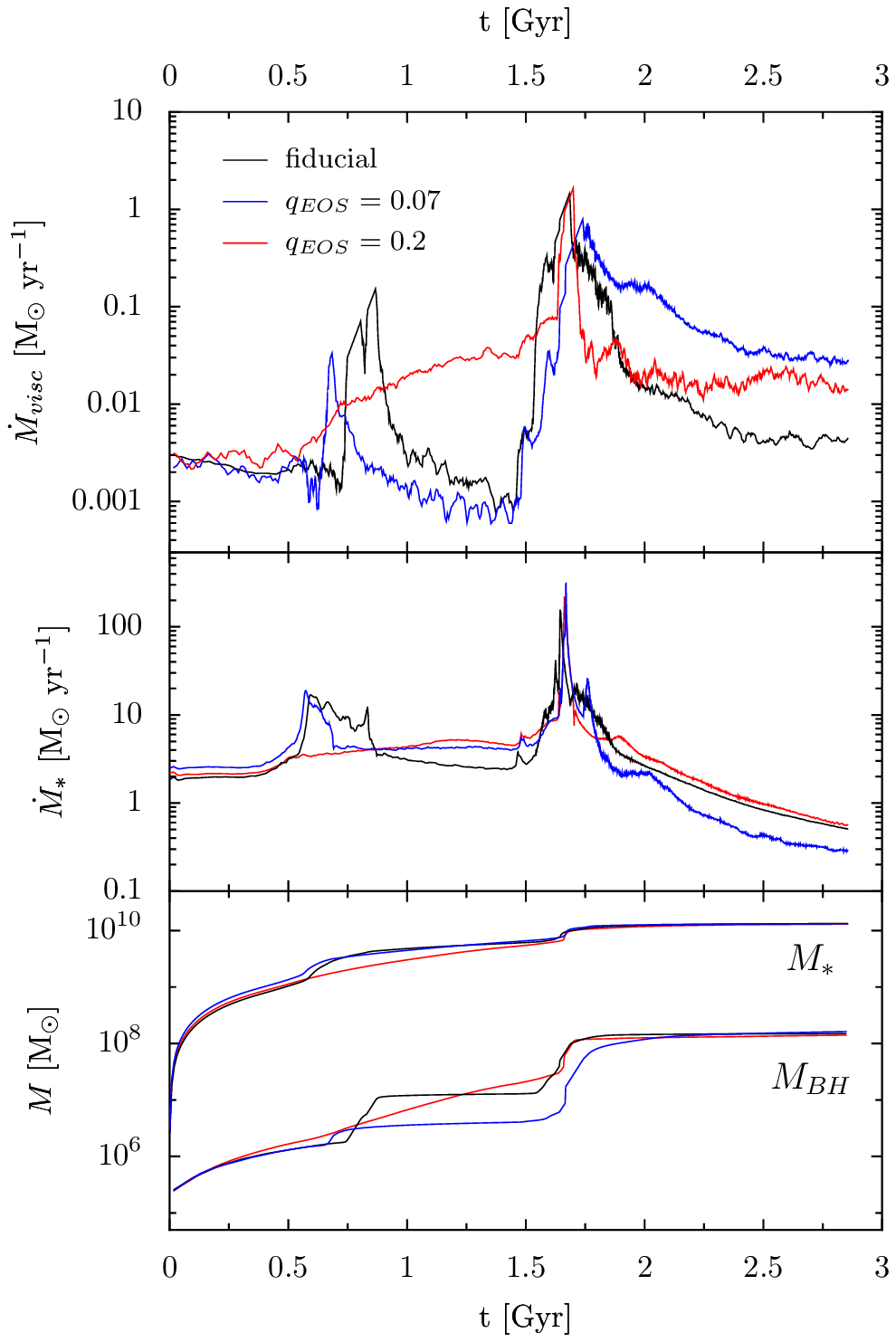}
\caption{Comparison of three simulations that differ only in the ISM
  models: fiducial (black), $q_{EOS} =0.2$ (red), and $q_{EOS} = 0.07$
  (blue).  The panels show the viscous accretion rate (top), star
  formation rate (middle), and the integrated black hole mass and mass
  of new stars formed (bottom).  The three different ISM models have
  $c_s$ and Toomre $Q$ within a factor of $\sim 2$ of one another at
  all radii; the $\qeos = 0.2$ model has the largest values of $c_s$
  and Q and our fiducial model has the smallest values. }
\label{figureISMSeries}
\end{figure}

Fig.~\ref{figureISMSeries} compares the BH accretion history (top
panel), the star formation rate (middle), and the integrated BH mass
and mass of new stars formed during the merger (bottom) for the three
runs with differing ISM models.  For both the fiducial run and the run
with $q_{EOS} = 0.07$ there is significant fragmentation after first
passage, which generates the first peak in star formation and BH
accretion.  By contrast, the run with $q_{EOS} = 0.2$ shows no
evidence for gas fragmentation or a pronounced peak in activity at
first passage.  Despite these differing initial histories, the final
result of the merger is very similar in all three cases: the large
star formation rates and BH accretion rates coincident with the final
coalescence of the two galaxies are not due to fragmentation, but are
instead largely due to the inflow of diffuse gas to smaller radii.
Moreover, the final BH mass and the total amount of new stars formed
during the merger are similar in all three cases.
Thus, despite uncertainties in the model
of the ISM, we find relatively robust integrated quantities (as did
the earlier calculations of \citealt{1995ApJ...448...41H}). The
precise time dependence of the star formation and BH accretion (i.e.,
the lightcurves) are, however, significantly more uncertain and
sensitive to the details of the model.

\vspace{-0.4cm}
\subsection{Galaxy Parameters}
\label{sec:gal}

Having shown that the final BH mass and new stellar mass do not depend
strongly on the uncertain parameters in our accretion, feedback and
ISM models, we now examine how our results vary with galaxy properties
such as the total mass, gas fraction, and bulge-to-disk ratio.

\begin{figure}
\includegraphics[width=84mm]{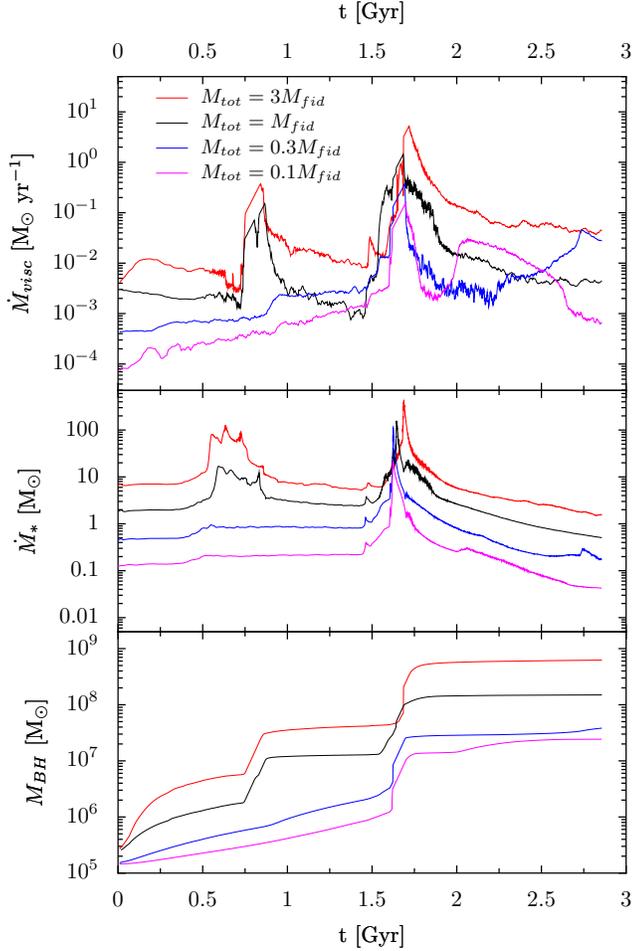}
\caption{Comparison of four simulations that differ only in the total
  galaxy mass: $M_{fid}=3.88\times 10^{12}M_\odot$ (fiducial; black), $3
  M_{fid}$ (red), $0.3 M_{fid}$ (blue), and $0.1 M_{fid}$ (magenta).  The
  three panels show the viscous accretion rate (top), star formation rate
  (middle), and the integrated black hole mass (bottom).  The same
  parameters are used in the BH accretion and feedback models.  }
\label{FigureMassScale}
\end{figure}  

Fig.~\ref{FigureMassScale} shows the BH accretion histories (top
panel), star formation rate (middle), and integrated BH mass (bottom)
for four runs with different total galaxy mass.  The models cover a
factor of 30 in galaxy mass, from 0.1-3 times our fiducial mass.  The
BH and star formation parameters are identical in the four
simulations, while the gravitational force softening and structural
parameters (e.g., disk scale length, bulge radius) change with the
total mass (see \S \ref{sectionICs}).

Fig.~\ref{FigureMassScale} shows that the BH accretion rates and
integrated BH masses increase with galaxy mass as expected from
equation~\ref{EquationMCrit}.  However, there is a clear difference
between the lower and higher mass simulations: the two higher mass
simulations show evidence for the first peak in star formation and BH
growth that we have shown is due to fragmentation near first passage,
while the lower mass runs do not.  This is largely a consequence of
the fact that observed disks have $R_D \propto M^{1/3}$
\citep{2003MNRAS.343..978S}, so that more massive galaxies have higher
surface densities and are thus more susceptible to gravitational
instability (our ISM model counteracts this slightly, but not enough
to stabilize the higher mass disks).  It is important to reiterate,
however, that modest changes to the subgrid sound speed can change
whether or not the gas fragments near first passage (\S \ref{sec:ISM})
so it is not clear if the difference as a function of mass in
Fig.~\ref{FigureMassScale} is robust.

In addition to the systematic change in the importance of
fragmentation near first passage, Fig.~\ref{FigureMassScale} also
shows differences in the late-time BH accretion between the low and
high mass simulations.  In particular the two smaller mass runs each
show a period of increased accretion after the main peak during the
merger.  In these cases the new stars formed around final coalescence
develop a bar in the inner $\sim R_{acc}$ of the galaxy. This helps
drive some of the remaining gas into the accretion region leading to
the increased accretion at late times.  There is a milder version of
this late-time accretion in the fiducial mass $\qeos = 0.2$ model
without fragmentation in Fig. \ref{figureISMSeries}.  Interestingly,
there is no analogous late-time inflow of gas to within $R_{acc}$ in
our low mass galaxy simulations without BH feedback.  The late-time
activity is also particularly sensitive to the accretion model at a
time when the non-axisymmetry produced by the merger has died away (so
that $\alpha$ may in reality decrease significantly).  For these
reasons, we regard the late time growth in
Fig.~\ref{FigureMassScale} as an interesting deviation from
self-similarity in the dynamics, but not necessarily a particularly
robust one.  One important point that this highlights, however, is
that because our implementation of BH growth and feedback does not
unbind a significant amount of cold gas at late times (unlike
calculations by \citealt{springel2005b}), the predictions of our
model are more sensitive to the post galaxy coalescence physics.

In addition to the fiducial gas fraction ($f_g = 0.1$) simulations
that we have largely focused on, we performed simulations with an
initial gas fraction of $f_g = 0.3$ for our fiducial galaxy mass and
at one tenth this mass.  The qualitative difference in behavior with
galaxy mass in Fig.~\ref{FigureMassScale} persists in the higher gas
fraction runs.  In particular, in the low mass $f_g = 0.3$ simulation,
the gas does not fragment, while it does in the higher mass $f_g =
0.3$ simulation. Fig. ~\ref{FigureSigmaFiducial} -- discussed in \S
\ref{sec:fid} -- explicitly shows the increase in the gas surface
density within $R_{acc}$ produced by this at early times.

A final property of the galaxy model that we varied was the bulge to
disk mass ratio.  The majority of our runs include a bulge with one
third the mass of the disk; we also ran one simulation with an initial
bulge of one fifth the disk mass, at the fiducial galaxy mass.  The
final BH mass and total mass of stars formed differ by less than $3\%$
each compared to the fiducial simulation.

\vspace{-0.5cm}
\section{The  $M_{BH}-\sigma$ Correlation}
\label{sectionGalaxyBH}

Previous numerical studies using models of BH growth and feedback different
from those considered here have reproduced a number of the observed
correlations between massive BHs and their host galaxies (e.g.,
\citealt{di-matteo2005,sazonov2005,younger08}). \citet{younger08} argue
that the galaxy-BH correlations in simulations (in particular, the BH
fundamental plane) are relatively independent of the trigger of BH growth,
be it minor mergers, major mergers, or global instabilities of galactic
disks.  Based on the calculations to date, however, it is unclear to what
extent the simulated BH-galaxy correlations depend on the details of the BH
feedback or accretion models.  In this section we assess this question by
quantifying the $M_{BH}$ - $\sigma$ relation produced in our models.

We define $\sigma$ of our model galaxies using a method analogous to
that of observers: we first project the mass density of the stellar
particles into cylindrical bins, and compute the half-mass(light)
radius $R_e$.  We then compute the velocity dispersion weighted by the
surface brightness via
\begin{equation}
\sigma^2 = \frac{\int_{R_{min}}^{R_e} \sigma_{los}^2(R) I(R) R dR}{\int_{R_{min}}^{R_e} I(R) R dR}
\label{EquationSigmaLOS}
\end{equation}
\noindent where $I(R)$ is the {projected 2-d stellar} mass profile,
$\sigma_{los}$ is the line of sight velocity dispersion, and $R_{min} = 2
\epsilon$ to ensure that there are that no artificial effects introduced by
the force softening.  We repeat this calculation along $1000$ lines of
sight with random viewing angles through the center of mass of the merger
remnant.  The $\sigma$ quoted in this paper and listed in Table~1 is the
median value over the 1000 lines of sight.

\begin{figure}
\includegraphics[width=86mm]{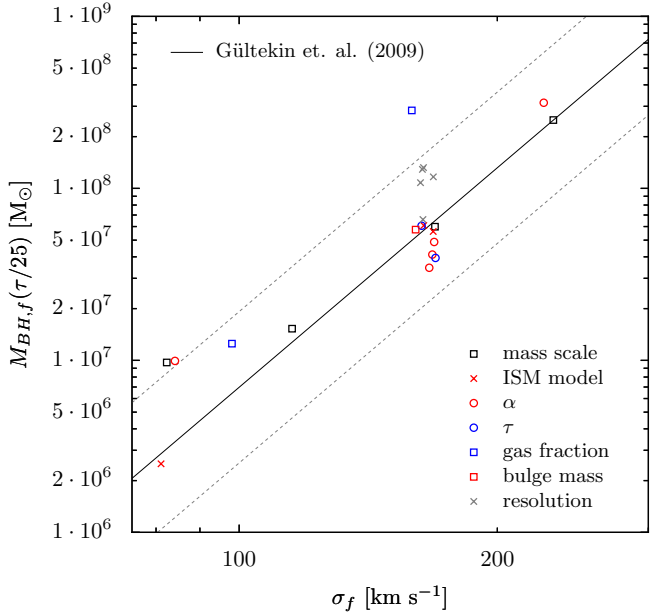}
\caption{The $M_{BH,f}$-$\sigma$ relation for the simulations in this
  paper, along with the observed relation (solid) and one sigma
  scatter (dotted) from \citet{2009ApJ...698..198G}.  The final BH
  mass $M_{BH,f}$ in all of the simulations has been linearly scaled
  to $\tau = 25$ from the value used in the simulation (typically
  $\tau = 10$). The simulations are generally quite consistent with
  observations; we do find indications of a slight flattening in
  $M_{BH,f}-\sigma$ at low BH masses.}
\label{FigureSigmaPlot}
\end{figure}

Fig.~\ref{FigureSigmaPlot} shows the correlation between the final BH
mass $M_{BH,f}$ and the $\sigma$ of the merged galaxy for most of the
simulations in Table~\ref{TableRunParam}: different total galaxy
masses (black), different values of the accretion parameter $\alpha$
(red circle), alternate ISM models (red x), higher gas fraction (blue
square), alternate bulge mass (red square), different values of $\tau$
(blue circle), and the resolution studies in Appendix A (grey).  The
solid line indicates the mean relation from the compilation of
observational results in \cite{2009ApJ...698..198G} while the dotted
lines are the $1-\sigma$ error bars.  We have linearly rescaled all of
our final BH masses to a value of $\tau = 25$, using the fact that
both the analytic and numerical results are consistent with $\dot
M_{visc}$ and $M_{BH,f}$ being $\propto \tau^{-1}$.  The value of
$\tau = 25$ is chosen so that the rescaled fiducial simulation lies
approximately on the $M_{BH}-\sigma$ relation of
\cite{2009ApJ...698..198G}.  For our fiducial simulation carried out
with $\tau = 3$ and $\tau = 10$, a linear scaling of $M_{BH,f}$ with
$\tau^{-1}$ is accurate to about $2 \%$ (e.g.,
Table~\ref{TableRunParam} and Fig.~\ref{FigureFourPanel}).  We also
carried out our fiducial simulation with $\tau = 25$; this is
consistent with a linear scaling of $M_{BH,f}$ from $\tau = 3$ to
$\sim 50 \%$ (Table~\ref{TableRunParam}).  For nearly all of our
simulations, rescaling to $\tau = 25$ amounts to dividing the final BH
mass by a factor of 2.5.

\begin{figure}
\includegraphics[width=86mm]{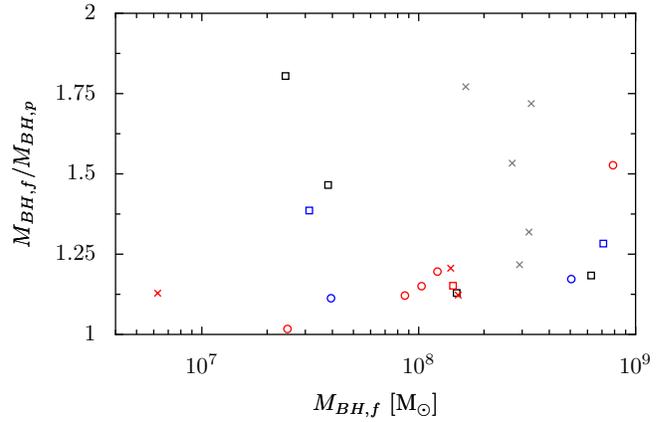}
\caption{The ratio of the final BH mass to the BH mass at the peak of
  accretion for the simulations in Fig.~\ref{FigureSigmaPlot}, using
  the same symbol types. This quantifies the extent to which late-time
  accretion increases the BH mass.  The late-time increase in BH mass
  for many of the lower mass systems produces the slight flattening in
  $M_{BH}-\sigma$ in Fig.  \ref{FigureSigmaPlot} at low masses; see
  the text for a discussion of the robustness of this result.}
\label{FigureSigmaPlotPeak}
\end{figure}

Previous analytic arguments were able to reproduce the $M_{BH}-\sigma$
relation with $\tau \sim 1$, rather than requiring $\tau \sim 25$ as
we do here (e.g., \citealt{king2003,murray2005}).  These calculations,
however, assumed $f_{g} = 0.1$.  While perhaps appropriate on average,
this is not appropriate in galactic nuclei where the gas densities are
higher.  The analytic derivations also assumed that BH growth
terminated when the system reached the luminosity (accretion rate) at
which radiation pressure balances gravity (eq.~\ref{EquationMCrit}).
In reality, however, the luminosity must exceed this critical value by
a factor of several in order for gas to be efficiently pushed around
in the galactic nucleus (as shown explicitly in the test problems in
the Appendix).  Moreover, the BH continues to accrete some mass even
after reaching $\dot M_c$.  Fig.~\ref{FigureSigmaPlotPeak} shows this
explicitly via the ratio of the final BH mass to the BH mass at the
peak of activity for all of the simulations in
Fig.~\ref{FigureSigmaPlot}.\footnote{To account for the fluctuating
  nature of the BH accretion rate in some of the simulations, we
  define the BH mass at ``peak'' to be the mass when $\dot M$ drops by
  a factor of 10 from its peak value.}  The net effect of the
differences between our simulations and the simple analytic
calculations is that a much larger feedback force per unit BH mass
($\tau \sim 25$, not $\sim 1$) is required for consistency with the
observed $M_{BH}-\sigma$ relation.  The physical implications of this
larger value of $\tau$ for models of AGN feedback will be discussed in
\S~\ref{sectionDiscussion}.

The scatter in BH mass in Fig.~\ref{FigureSigmaPlot} at our fiducial
mass scale of $\sigma \sim 175 \kms$ is reasonably consistent with the
observed scatter. In the simulations we have varied the BH accretion
model ($\alpha$), the ISM model, numerical resolution, size of the
feedback/accretion region $R_{acc}$, and galaxy properties such as the
total mass, gas fraction, and bulge to disk ratio.  It is encouraging
that all of these simulations produce BH masses within a factor of few
of each other. The largest BH mass at $\sigma \sim 175 \kms$ is the
simulation with an initial gas fraction of $f_g = 0.3$; since this run
has a larger gas density at small radii close to the BH
(Fig.~\ref{FigureSigmaFiducial}), it should probably also have a
larger $\tau$, which would reduce the BH mass further, in better
agreement with the data.  It is difficult to make this comparison to
the observed scatter more quantitative given the limitation that our
simulations are all equal-mass non-cosmological binary mergers on the
same orbit.

The numerical results in Fig.~\ref{FigureSigmaPlot} suggest a slight
flattening of the $M_{BH}-\sigma$ relation at $\sigma \lesssim 100
\kms$.  This is in large part a consequence of the additional mass
gained by the lower mass BHs after their peak of activity (see Figs.
\ref{FigureMassScale} \& \ref{FigureSigmaPlotPeak}, in particular the
fiducial simulations labeled by black squares in
Fig.~\ref{FigureSigmaPlotPeak}).  This change in behavior at lower
masses is primarily due to the fact that the lower mass galaxies are
less prone to fragmentation than the more massive galaxies (\S
\ref{sec:gal}).  Without the fragmentation after first passage, more
gas is available to feed the BH at late times leading to the slightly
higher BH mass.  As discussed in \S \ref{sec:gal}, it is unclear how
robust this late time accretion is.  In fact, a low mass galaxy
simulation with an alternate ISM model ($\qeos = 0.07$) does not show
significant late-time accretion, leading to a BH mass in good
agreement with the extrapolation from higher $\sigma$ (red x at low
mass in Figs. \ref{FigureSigmaPlot} \& \ref{FigureSigmaPlotPeak}).  We
thus regard the case for flattening of $M_{BH}-\sigma$ at low masses
in our models as somewhat tentative; our results may instead indicate
enhanced scatter at low masses rather than a change in the mean
relation.  More comprehensive numerical studies of these lower mass
systems will be needed to distinguish these two possibilities.

\vspace{-0.5cm}
\section{Discussion and Conclusions}
\label{sectionDiscussion}

We have presented a new method for simulating the growth of massive
BHs in galaxies and the impact of AGN activity on gas in its host
galaxy (see also our related Letter; \citealt{2009arXiv0909.2872D}).
In this method, we use a local viscous estimate to determine the
accretion rate onto a BH given conditions in the surrounding galaxy
(eq. \ref{mdotvisceqn}), and we model the effect of BH feedback on
ambient gas by depositing momentum radially away from the BH into the
surrounding gas (eq.~\ref{momdepeqn}).

Our accretion model qualitatively takes into account the angular momentum
redistribution required for accretion of cold gas in galaxies and is thus
more appropriate than the spherical accretion estimate that has been used
extensively in the literature.  In our feedback model, the applied force is
given by $\tau L/c$, where the AGN's luminosity $L$ is determined by our BH
accretion model, and the net efficiency of the feedback is determined by
the total optical depth $\tau$ of the galactic gas to the AGN's radiation,
which is a free parameter of our model.  Previous calculations have
demonstrated that only when the gas fraction in a galaxy decreases to
$\lesssim 0.01$ can the AGN's radiation Compton heat matter to high
temperatures \citep{sazonov2005}.  More generally, the cooling times in
gas-rich galaxies are so short that the primary {dynamical} impact of the
AGN on surrounding gas is via the momentum imparted by the AGN's outflows
or radiation.  It is thus not physically well-motivated to model AGN
feedback by depositing energy, but not momentum, into surrounding gas, as
many calculations have done (e.g., \citealt{di-matteo2005,
  springel2005b,kawata2005}); see \cite{ostriker10b} for related points.

Throughout this paper, we have focused on BH growth during major
mergers of spiral galaxies.  As demonstrated in
\citet{2009arXiv0909.2872D}, our model leads to a self-regulated mode
of BH accretion in which the BH accretion rate is relatively
independent of the details of the BH accretion model (see
Fig.~\ref{FigureFourPanel}).  This is because the accretion rate
self-adjusts so that the radiation pressure force is comparable to the
inward gravitational force produced by the host galaxy (see
eq.~\ref{EquationMCrit}).  This self-regulated mode of BH
accretion is a robust feature of all of our simulations during periods
of time when there is a significant nuclear gas reservoir -- it thus
applies precisely when the BH gains most of its mass.

One important consequence of this self-regulated accretion is that AGN
feedback does not drive significant large-scale outflows of gas (in
contrast to the models of \citealt{springel2005b}).  For example, the
surface density profiles in Fig.~\ref{FigureSigmaofRFiducial} show
that AGN feedback causes gas to pile up at a few hundred pc rather
than being completely unbound from the galaxy -- this precise radius
should not be taken too literally since it is a direct consequence of
the fact that we implement feedback and determine the BH accretion
rate only within a radius $R_{acc} \sim$ few hundred pc.  Nonetheless,
we believe that this general result may well be generic: because the
BH accretion rate is determined by the gas content close to the BH,
the AGN can shut off its own accretion before depositing sufficient
energy to unbind all of the gas in the galaxy.  If we artificially
hold the luminosity of the AGN constant in time at a value exceeding
the critical value in equation~(\ref{EquationMCrit}), then the AGN {\em
  does} eventually unbind all of the surrounding gas (see, e.g.,
Figs.~\ref{FigureShellTestEta} \& \ref{FigIsothermalEta00} in Appendix
B).  However, both our isothermal sphere test problem
(Fig. \ref{FigureShellFull}) and our full merger calculations show
that when the BH accretion rate is self-consistently determined by the
gas properties in the central $\sim 100$ pc of the galaxy, the AGN
simply never stays `on' long enough to unbind all the gas.

Our results do not, of course, preclude that AGN drive galactic winds.
For example, some gas may be unbound by a high speed wind/jet produced
by the central accretion disk (which is not in our simulations).  In
addition, at later stages of a merger or at large radii the gas
fraction can be sufficiently low ($\lesssim 0.01$) that gas can be
Compton heated to high temperatures and potentially unbound (e.g.,
\citealt{ciotti10}).  This may in fact be sufficient to quench star
formation at late times, but only once most of the gas has already
been consumed into stars (so that $f_g \lesssim 0.01$).  Our results
do suggest that AGN feedback does not quench star formation by
unbinding a significant fraction of the cold dense gas in a galaxies
interstellar medium (in contrast to, e.g., \citealt{springel2005b}).
In future work it will be important to assess whether variability in
the accretion rate on smaller scales than we can resolve (e.g.,
\citealt{hq10,levine10}) modifies this conclusion; such variability
could produce some epochs during which AGN feedback has a
significantly larger effect on the surrounding gas.  Another
improvement would be to carry out radiative transfer
calculations and assess what fraction of the AGN's radiation is
absorbed at large radii in a galaxy ($\sim$ kpc) where the gas has a
lower surface density and is thus easier to unbind.

Our simulations cover a factor of $\sim 30$ in galaxy mass. The final
BH mass in our calculations is $\propto \tau^{-1}$ since a larger
value of $\tau$ corresponds to a larger momentum deposition per unit
BH mass.  We find reasonable consistency with the normalization of the
observed $M_{BH}-\sigma$ relation for $\tau \sim 25$.  To compare this result
to previous work by \citet{di-matteo2005}, we note that a momentum
deposition of $\dot P$ corresponds to an energy deposition rate of
$\dot E \simeq \dot P \sigma$ when the feedback is able to move the
gas at a speed comparable to the velocity dispersion $\sigma$ (which
is required for efficient self-regulation of the BH growth).  For
$\tau \simeq 25$, our results thus correspond to $\dot E \simeq 25 \,
L \, \sigma/c \simeq 0.02 \, L \, (\sigma/200 \, {\rm km \, s^{-1}})$.
This is similar to the results of \citet{di-matteo2005}, who found
that depositing $\sim 5 \%$ of the BH accretion energy in the
surrounding gas was required to explain the $M_{BH}-\sigma$ relation.
It is encouraging that these two different sets of simulations, with
different BH accretion and feedback models, 
agree at the factor of $\sim 2-3$ level on the energetics
required to reproduce the $M_{BH}-\sigma$ relation.

The value of $\tau \sim 25$ required to explain the normalization of
the $M_{BH}-\sigma$ relation has strong implications for the dominant
physics regulating BH growth.  The simplest models of super-Eddington
winds from an accretion disk close to the BH are ruled out because
they typically have $\tau \sim 1$, i.e., a momentum flux comparable to
that of the initial radiation field \citep{king2003}.  Similarly, the
radiation pressure force produced solely by the scattering and
absorption of the AGN's UV radiation by dust corresponds to $\tau \sim
1$ \citep{murray2005} and is thus not sufficient to account for the
level of feedback required here.  Rather, our results suggest that
most BH growth happens when the nuclear regions are optically thick to
the re-radiated dust emission in the near and far-infrared, so that
$\tau \gg 1$.  This is consistent with observational evidence in favor
of a connection between BH growth, quasars and luminous
dust-enshrouded starbursts such as ULIRGs and sub-mm galaxies (e.g.,
\citealt{1988ApJ...325...74S,dasyra2006,alexander2008}).
Quantitatively, the observed {\it stellar} densities at radii $\sim
1-100$ pc in elliptical galaxies reach $\sim 20$ g cm$^{-2}$
\citep{hopkins10b}, implying $\tau \sim 100$ if a significant fraction
of the stars were formed in a single gas-rich epoch.  It is
encouraging that this is within an order of magnitude of (and larger
than!) the value of $\tau$ we find is required to explain the observed
$M_{\rm BH}-\sigma$ relation.

A fixed value of $\tau \sim 25$ independent of galaxy mass produces an
$M_{BH}-\sigma$ relation with a slope and scatter in reasonable
agreement with observations (see Fig.~\ref{FigureSigmaPlot}).
Assessing the scatter more quantitatively will require a wider survey
of merger orbits.  We do find some tentative evidence for a shallower
slope in the $M_{BH}-\sigma$ relation at the lowest galaxy masses,
corresponding to $\sigma \lesssim 100 \kms$.  This range of masses is
precisely where the observational situation is particularly unclear,
with, e.g., possible differences between the BH-galaxy correlations in
classical bulges and pseudo-bulges \citep{greene08}.  It is also
unclear whether major mergers are the dominant mechanism for BH growth
in these lower mass galaxies (e.g., \citealt{younger08}).

Our simulations show that fragmentation of a galactic disk into clumps
can be efficiently {\em induced} by a merger (e.g.,
Fig. \ref{figureSFRClump}), even when an isolated galaxy with same
properties is Toomre stable (see, e.g., \citealt{2007MNRAS.375..805W} for
related ideas in the context of dwarf galaxy formation in tidal
tails). As Figure \ref{figureISMSeries} demonstrates, this
fragmentation can produce a significant increase in star formation
during the first close passage of galaxies even when there is little
inflow of the diffuse ISM (because such inflow is suppressed by a
bulge until later in the merger; \citealt{1996ApJ...464..641M}). In
our simulations we often see a corresponding increase in the BH
accretion rate due to the inspiral of dense gas-rich clumps
(Fig. \ref{figureSFRClump}).  The inflow of {gas} by this process may,
however, be overestimated: stellar feedback not included in our
simulations can unbind the gas in star clusters on a timescale of
$\sim$ a Myr, returning most of the gas to the diffuse ISM (e.g.,
\citealt{murray10,hq10}).

Our calculations use subgrid sound speeds motivated by the
observed turbulent velocities in galaxies (\S \ref{sectionSFR}).  We
thus believe that our ISM model is physically well-motivated, even
though the use of a subgrid sound speed necessarily introduces some
uncertainty.  Overall, the presence/absence of large-scale clumping of
the ISM does not significantly change the final BH mass or the mass of
new stars formed in our simulations. It can, however, change the star
formation rate and BH accretion rate as a function of time,
particularly near the first close passage during a merger.

The tentative change in the $M_{BH}-\sigma$ relation we find for lower
mass galaxies is largely due to our treatment of the ISM, rather than
our BH feedback or accretion model.  For a given gas fraction, lower
mass galaxies have a lower gas surface density and thus the ISM is
less prone to fragmentation (\S \ref{sec:gal} and
Fig. \ref{FigureMassScale}).  Without the fragmentation after first
passage, more gas is available to feed the BH at late times leading to
somewhat higher BH mass (Fig.~\ref{FigureSigmaPlotPeak}).

The BH accretion and feedback models used in this paper can be
significantly improved in future work, allowing a more detailed
comparison to observations.  For example, \citet{hq10} carried out a
large number of simulations of gas inflow in galactic nuclei from
$\sim 100$ pc to $\lesssim 0.1$ pc (see, e.g., \citealt{levine10} for
related work).  These can be used to provide a more accurate estimate
of the BH accretion rate given conditions at larger radii in a galaxy
(Hopkins \& Quataert, in prep).  Another important improvement would
be to use a radiative transfer calculation to self-consistently
determine the infrared radiation field produced by a central AGN (and
distributed star formation).  This could then be used to calculate the
radiation pressure force on surrounding gas, eliminating the need for
our parameterization of the force in terms of the optical depth
$\tau$.

\vspace{-0.15cm}
\section*{Acknowledgments}

We thank Phil Hopkins and Yuval Birnboim for useful conversations. JD
and EQ were supported in part by NASA grant NNG06GI68G and the David
and Lucile Packard Foundation. Support for EQ was also provided in
part by the Miller Institute for Basic Research in Science, University
of California Berkeley.  This research used resources of the National
Energy Research Scientific Computing Center, which is supported by the
Office of Science of the U.S.  Department of Energy under Contract
No. DE-AC02-05CH11231.  This work was partially supported by the
National Center for Supercomputing Applications under AST080048 and
utilized the Intel 64 cluster Abe. The authors acknowledge the Texas
Advanced Computing Center (TACC) at The University of Texas at Austin
for providing HPC resources that have contributed to the research
results reported within this paper.

\bibliographystyle{mn2e}
\bibliography{paper_bib,main_bib}

\appendix

\vspace{-0.4cm}
\section{Resolution Studies}   
\label{resolution}

\begin{figure*}
\includegraphics[width=180mm]{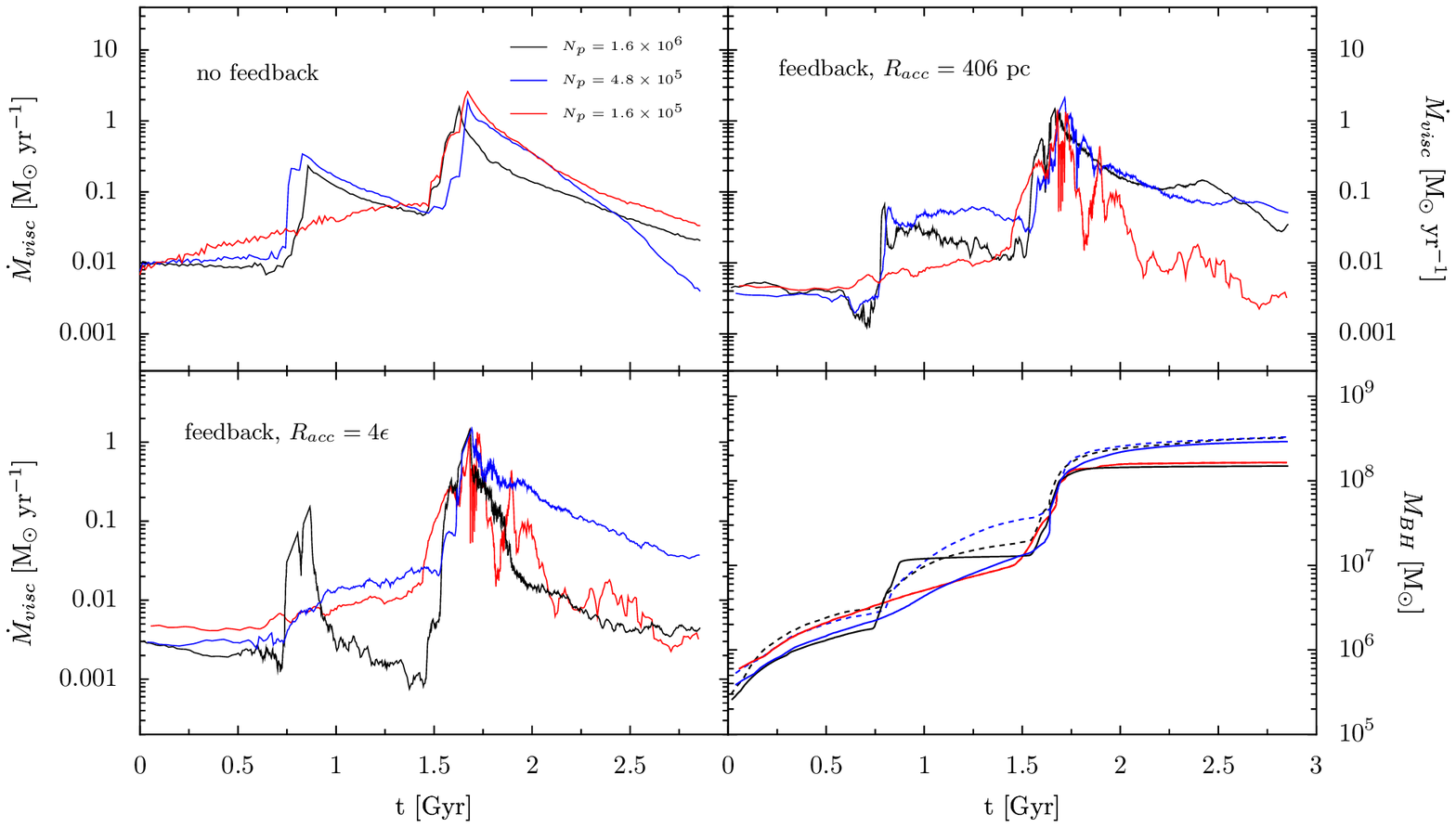}
\caption{\emph{Top left:} BH accretion rate for simulations without
  feedback at three resolutions (LRfidNof, MRfidNof, fidNof in red,
  blue and black respectively); $\dot M_{visc}$ was computed for
  $\alpha = 0.05$ and using the same value of $R_{acc} = 406$ pc at
  all three resolutions (this corresponds to $4 \epsilon$ for the
  lowest resolution).  \emph{Top right:} BH accretion rate with
  feedback at the same three resolutions using $R_{acc} = 406$ pc.
  \emph{Bottom left:} BH accretion rate with feedback at the same
  three resolutions using $R_{acc} = 4 \epsilon$; here the accretion
  rate and feedback are calculated in different physical volumes at
  different resolutions.  \emph{Bottom right:} BH masses for all of
  the runs in this Figure with feedback.  Solid lines are for $R_{acc}
  = 4\epsilon$ while dashed lines are for $R_{acc} = 406$ pc.}
\label{FigureResoFourPanel}
\end{figure*}

In this section, we describe some of our resolution tests both with
and without BH feedback.  In the absence of feedback, the well-posed
questions for resolution studies include both how the gas properties
as a function of radius and time depend on the resolution and how
integrated properties of the galaxy (e.g., the star formation rate)
depend on resolution.  However, the feedback, when present, has a
nontrivial dependence on the resolution and it is by no means clear
that the nonlinear system will in fact converge in a simple way with
increasing resolution.  Physically, e.g., the AGN's radiation pressure
has the strongest effect on the gas that contributes the most to the
optical depth, which is largely determined by the column density (the
dust opacity being only a relatively weak function of temperature for
the conditions of interest).  Higher resolution simulations can
resolve higher volume and column densities, largely at smaller radii
close to the BH, and thus may change some of the details of the BH
feedback.  Indeed, Fig. \ref{FigureSigmaofRFiducial} shows that the
column density increases towards smaller radii in our simulations.

We first consider the question of how the nuclear gas properties
depend on numerical resolution \emph{in the absence of feedback}.  To
this end, the top left panel of Fig.~\ref{FigureResoFourPanel} shows
the BH accretion rate $\dot M_{visc}$ calculated for three different
particle numbers $N_p = 1.6 \times 10^5, 4.8 \times 10^5,$ and $1.6
\times 10^6$, with the gravitational force softening $\epsilon \propto
N_p^{1/3}$.\footnote{In Fig.~\ref{FigureResoFourPanel}, $\dot
  M_{visc}$ for the simulations without feedback (upper left) is
  calculated from the simulation snapshots and the accretion rate is
  not Eddington limited.  The data outputs were relatively infrequent
  and attempting to integrate the BH mass over such large timesteps
  was inaccurate.}  To make a fair comparison, the accretion rate is
evaluated within a fixed volume ($R = 406$ pc) and for $\alpha = 0.05$
for all of the simulations.  This choice corresponds to $R = 4
\epsilon$ for the lowest resolution run, but is $R \simeq 8.6
\epsilon$ for our fiducial resolution simulation.  Fig.
\ref{FigureResoFourPanel} shows that the lowest resolution simulation
(red) does not adequately resolve the fragmentation of the gas, and
the resulting peak in the accretion rate, near first passage.  The
medium and higher (= our fiducial) resolution simulations, however,
agree reasonably well, except for a slight difference in the slope of
$\dot M_{visc}(t)$ at late times.  Computed over a larger volume
($\sim$ kpc), the agreement between these runs improves.

To assess the convergence in the presence of feedback, the top right
panel of Fig.~\ref{FigureResoFourPanel} shows the BH accretion 
rate $\dot M_{visc}$ evaluated just as in the top left panel, 
i.e., using a fixed $R_{acc} = 406$ pc, in simulations
with the same three particle numbers and force softening.  Again the
lowest resolution (red) simulation is clearly not adequate, but the
medium (blue) and high (black) resolution simulations agree well; the
integrated BH mass differs only by 2\% in the latter two simulations.

As a final resolution test, the bottom left panel of
Fig.~\ref{FigureResoFourPanel} shows the BH accretion rate as a
function of time in simulations with the same three resolutions and
force softening, but in which $R_{acc} = 4 \epsilon$.  Thus in this
case the accretion rate is determined, and the feedback applied, on
increasingly small spatial scales in the higher resolution
simulations.  This is probably the most physically realistic (see \S
\ref{sectionParam}).  This panel shows that the large peak of
accretion at final coalescence ($t \sim 1.8$ Gyr) is quite similar in
all three cases.  This is set by the physics of feedback by momentum
deposition and is a robust property of all of our simulations.  A
corollary of this is that the final BH mass, as shown in the bottom
right panel of Fig.~\ref{FigureResoFourPanel}, is the same to within a
factor of $\sim 2$ for the three different resolutions. However, the
results in the lower left panel of Fig.~\ref{FigureResoFourPanel} also
clearly demonstrate that the detailed evolution of the accretion rate
is sensitive to the resolution. This is not particularly surprising:
at fixed resolution, Fig. \ref{FigureFourPanel} has already
demonstrated that the details of $\dot M_{visc}(t)$ depend on the
value of $R_{acc}$ -- although, again, neither the integrated BH mass
or star formation rate do.  One implication of these results is that
it is difficult for current simulations of BH growth to make
quantitative predictions about the light curves of AGN activity
triggered by mergers.

\vspace{-0.4cm}
\section{Code Verification}
\label{sectionAppendix}

We have tested our modifications to GADGET on a number of simplified
problems that have answers that can be easily obtained through other
methods.  \S B1 describes tests of the additional momentum feedback
force applied to a thin spherical shell of gas. \S B2 describes tests
in which the force is applied to the gas particles in the central
regions of an isothermal sphere.  Two ways of implementing the force
are tested: to a fixed number of particles around the BH, and to all
particles within a fixed region $R_{acc}$ around the BH.

As we are concerned with the performance of our BH accretion and
feedback model, in all of the tests presented in this appendix, the
multiphase equation of state and star formation model of
\cite{2003MNRAS.339..289S} are \emph{not} used; instead we use an
adiabatic equation of state with $\gamma = 5/3$.

\vspace{-0.3cm}
\subsection{Gas shells}

To test that the code is applying the radiation pressure force in 
equation~(\ref{momdepeqn}) correctly, we have run the code for a simple 
system containing a black hole particle with a large mass and a thin 
spherical shell of gas with negligible temperature, pressure and mass.  
As this gas resides in a thin shell, this problem is more well-posed if 
we apply the radiation force to a fixed number, $N$, of gas particles.

This system has a critical luminosity defined by the point at which the
radiation force balances the inward pull of gravity.  As the gas shell is
of low temperature and pressure, we can ignore pressure forces.  For a 
black hole of mass $M_{BH}$ and a gas shell of mass $m$ at a radius 
$r_0$ the critical luminosity $L_C$ satisfies (we take $\tau = 1$ for 
simplicity)

\begin{equation}
L_C = G \frac{M_{BH} m}{r_0^2} c.
\label{ShellCritLumo}
\end{equation}

\noindent When the luminosity is set to this value, the gas shell should 
experience no net force.  For other values of the luminosity, the 
expected behaviour can easily be calculated by noting that the gas shell, 
in the absence of any pressure forces, should have a radius, $r(t)$, that 
satisfies

\begin{equation}
m\frac{d^2 r(t)}{dt^2} = -\frac{G M_{BH} m}{r(t)^2} + \frac{L}{c}.
\label{ShellDiffEq}
\end{equation}

\noindent This is easily integrated to give the expected behavior of the
gas shell.

\begin{figure}
\includegraphics[width=84mm]{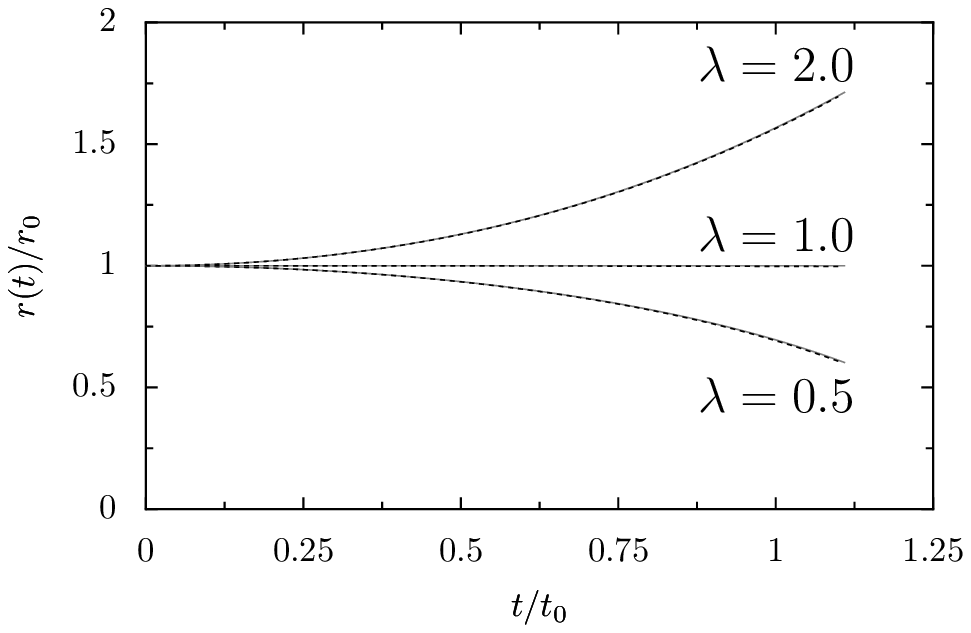}
\caption{Time evolution of the radius of the test shell for three
  values of radiation force: $\lambda = 0.5, 1.0, 2.0$ (dashed
  curves).  The results match closely with the solutions from
  integrating eq.~(\ref{ShellDiffEq}) (superposed grey curves).  Here
  the force is applied to the 25000 innermost gas particles of the
  $5\times10^4$ that make up the shell.  Time is in units of $t_0 =
  \sqrt{r_0^3/G M_{BH}}$ and the radius is in units of $r_0$, where
  $r_0$ is the initial radius of the gas shell.}
\label{FigureShellTestEta}
\end{figure}

A number of tests of this system were performed with varying luminosities, 
parameterized by the ratio of the luminosity applied to the critical 
luminosity, 

\begin{equation}
\lambda = \frac{L}{L_C}.  
\end{equation}

\noindent Fig.~\ref{FigureShellTestEta} shows the exact result in grey,
with the numerical solution from the modified version of GADGET in black,
for runs with $\lambda = 0.5, 1.0 \mbox{ and } 2.0$. For these tests the
number of particles in the shell is $N_{shell} = 50000$, and the force 
was applied to $N = 25000$ of them. In all cases, the numerical solution 
appears indistinguishable from the exact solution of 
eq.~(\ref{ShellDiffEq}).
 
\begin{figure}
\includegraphics[width=84mm]{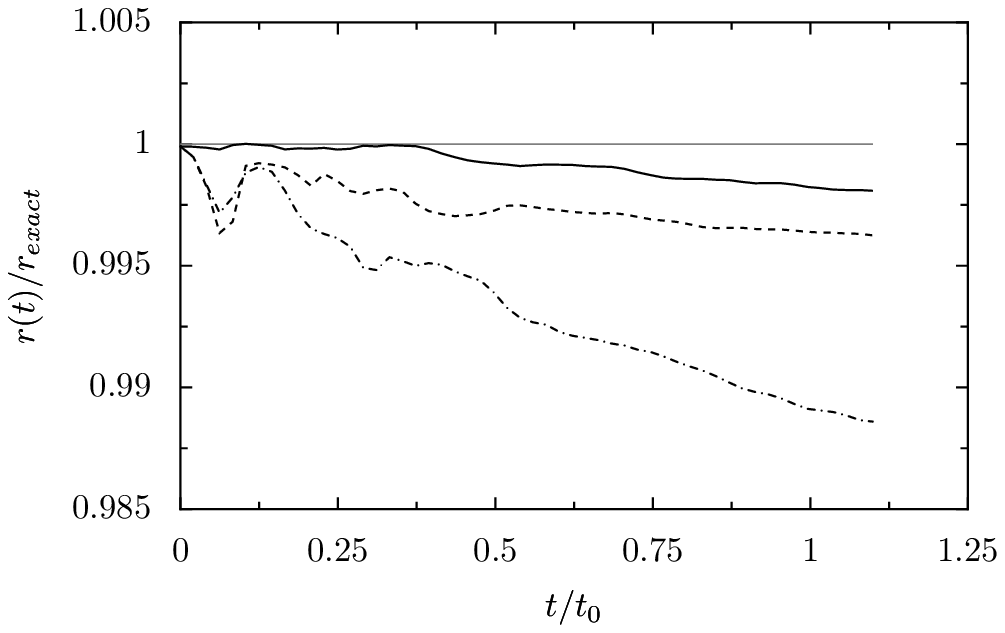}
\caption{Time evolution of the radius of the test shell for three
  values of $N / N_{shell}$: 0.5 (solid), 0.25 (dashed), and 0.1
  (dot-dashed).  The numerical solutions are normalized by the exact
  solution from eq.~(\ref{ShellDiffEq}).  The radiation force is fixed
  to be $\lambda = 2.0$.  The radius $r(t)$ changes by only about 1\%
  as $N$ is changed, indicating that our results are insensitive to
  the exact number of particles to which the radiation force is
  applied.}
\label{FigureShellTestXis}
\end{figure}

We have also tested the dependence of the results on the value of
$N/N_{shell}$, the fraction of particles that receives the radiation
force.  Fig.~\ref{FigureShellTestXis} shows the ratio of the
numerical solution from our code to the exact solution for $N /
N_{shell}= 0.5, 0.25$, and 0.1 for the $\lambda = 2.0$ model.  This
demonstrates that even though the magnitude of the force on an
individual particle increases as $N$ decreases, the overall dynamics
of the shell is the same, with the radii differing by only $\sim 1$\%
in the three cases.  This is primarily due to the fact that the SPH
particles are collisional and can thus transfer their motion to their
neighbors via pressure forces.  The extra momentum imparted to the
subset of particles is transferred in part to the outer region of the
shell, leading to the overall motion that agrees well with the exact
solution.  By extension, if $N$ were to vary over the duration of the
simulation, the results would also not depend strongly on the
particular value.

\vspace{-0.25cm}
\subsection{Isothermal Sphere}

We have performed a second set of tests of the feedback model using an
isothermal background given by a King model. The mass of the system is 
split into two parts.  The bulk of the mass makes up the collisionless 
background that is drawn from the full phase space distribution of the 
King model.  A small fraction of the mass, $f_g=0.05$, is assigned as 
collisional SPH particles.  These gas particles follow the same spatial 
profile as the collisionless background but are given zero initial 
velocities and a very low temperature.  Both components are realized 
with $10^5$ particles.  Finally, a black hole particle with a small mass 
is placed at rest at the center of the distribution.

\begin{figure*}
\includegraphics[width=180mm]{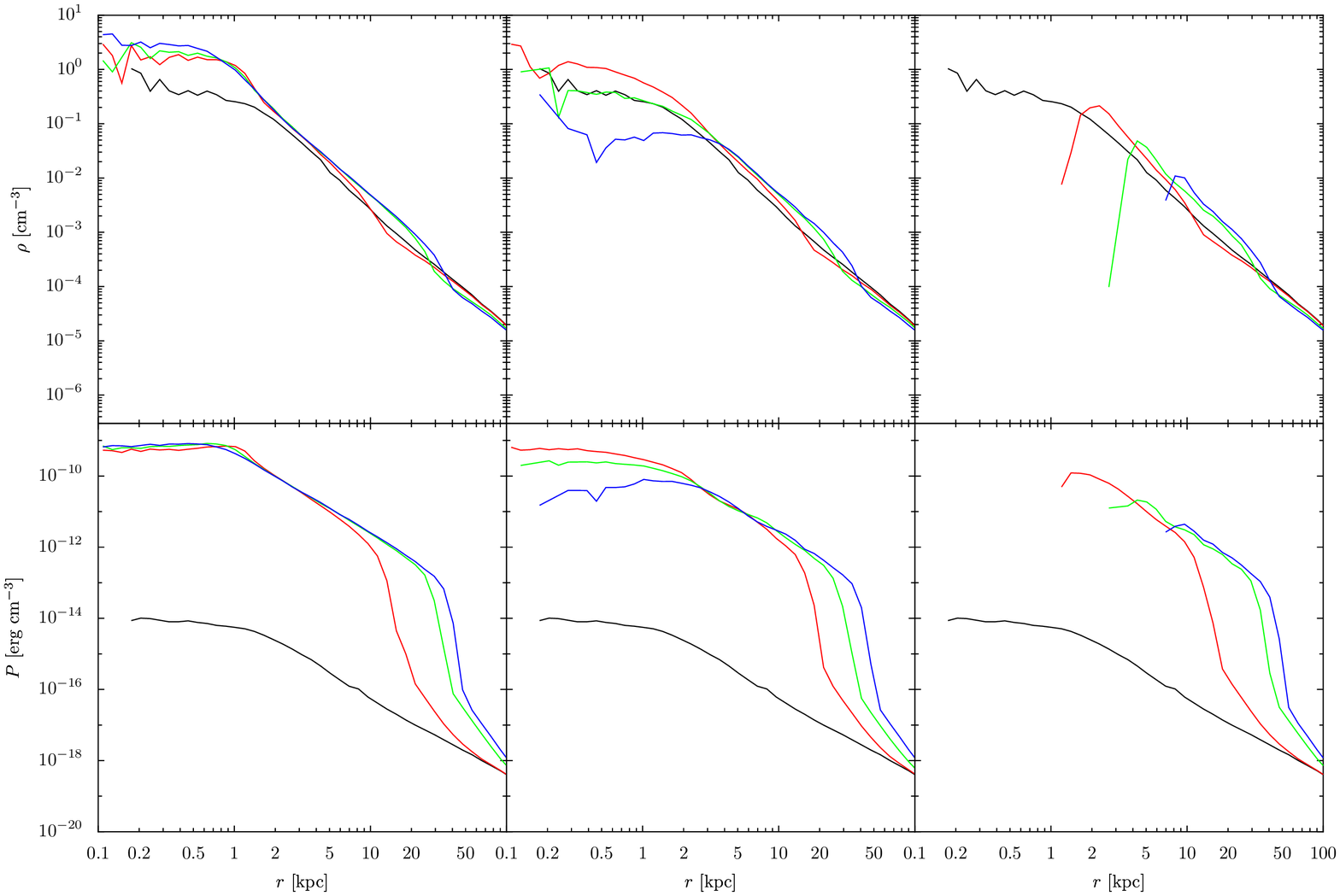}
\caption{The density (top) and pressure profiles (bottom) in
  simulations of an isothermal sphere of gas embedded in an isothermal
  King potential.  Three different simulations are shown: without
  feedback from a central black hole, $\lambda = 0.0$ (left), $\lambda
  = 1.0$ (middle) and $\lambda = 2.0$ (right).  Each simulation is
  shown at four times: the initial profile (black) and $t = 0.16$
  (red), 0.32 (green), 0.48 (blue) Gyr.}
\label{FigIsothermalEta00}
\end{figure*}

In the absence of feedback, the SPH particles are not in equilibrium by 
construction and should flow toward the center of the potential provided by
the collisionless background. When the feedback is switched on in the 
isothermal King potential near the center, the feedback will again have 
a critical value set by force balance:
\begin{equation}
\frac{L_c}{c} = 4 \frac{f_g \sigma^4}{G}.
\label{IsoThermCritLEqn}
\end{equation}
\noindent When the luminosity is below this value, we expect the extra
momentum to be insufficient to clear the gas out of the center.  When the
luminosity exceeds this value, the feedback should be strong enough to
clear the central regions of the distribution.  To test this, we apply
feedback with a constant luminosity.  Again, we parameterize the strength
of the feedback as $\lambda = L / L_c$.

We have tested two ways of assigning the radiation force.  In the first
case, the force is shared (equally) by a fixed number of gas particles
nearest to the black hole.  In the second case, the force is shared by all
gas particles within a fixed radius of the black hole.  We discuss the
results separately below.

\vspace{-0.2cm}
\subsubsection{Fixed $N$}

For the tests in this subsection, the radiation force is applied to a
fixed number of gas particles: $N=500$.  The King model has $\sigma =
100 \rm{km} \rm{s}^{-1}$, $\Psi/\sigma^2 = 12$ and a total mass of
$10^{12} M_{\sun}$.

Fig.~\ref{FigIsothermalEta00} compare the density and pressure profiles
of three runs with $\lambda=0$ (i.e. no feedback; left panels), 1 (middle),
and 2 (right).  Four timesteps are shown: $t=0$ (black), 0.16 (red), 
0.32 (green), and 0.48 Gyr (blue). As expected, the gas flows to the 
center in the absence of feedback, increasing the density and pressure 
as the gas begins to equilibrate in the background potential.  The middle 
and right panels show that the feedback clearly has an effect on the gas 
at the center, providing some support for the incoming gas, allowing the 
gas to have a lower pressure.  For the case with $\lambda=2$, the 
feedback is strong enough to effectively clear out the central region.

The nature of the feedback allows a calculation of how the size of the
evacuated region should grow with time.  Ignoring the thickness of the
shell swept up as matter begins to be driven out by the feedback,
momentum conservation gives

\begin{equation}
\frac{d}{dt}\left[M_{shell}(r) dr/dt \right] = \frac{L}{c} - \frac{G M_{bg}(r) M_{shell}(r)}{r^2}
\label{SnowPlowEqn}
\end{equation}

\noindent where $M_{shell}(r)$ is the initial mass distribution of gas and
$M_{bg}(r)$ is the mass distribution of the background.  Near the center of
the initial distribution, both the gas and background have an isothermal
distribution, for which the mass increases linearly with the distance from
the centre.  This makes the right hand side of Eq.~(\ref{SnowPlowEqn}) a
constant.  In this case, the size of the evacuated region, $r(t)$, depends
linearly on  time:
\begin{equation}
r(t) = \sqrt{2 (\lambda - 1) (1 - f_g)} \sigma t + C
\label{SnowPlowSoln}
\end{equation}

\noindent where $C$ is a constant of integration to account for the finite
time required to form the shell of swept up gas.

\begin{figure}
\includegraphics[width=84mm]{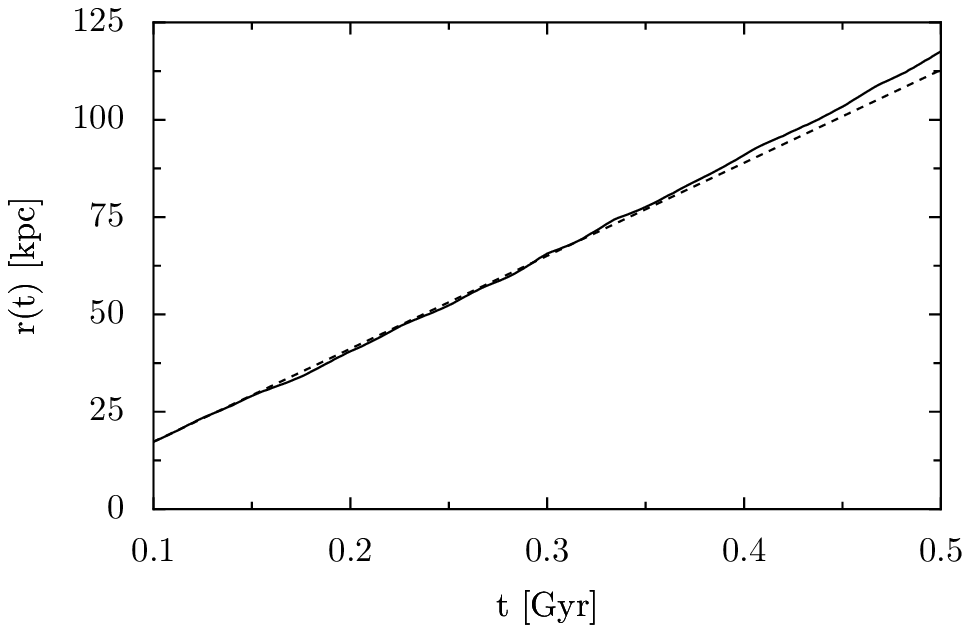}
\caption{Time evolution of the size of the evacuated region for the
  isothermal sphere test.  The $\lambda = 4$ simulation results shown
  (solid) match well with the analytic solution
  (Eq.~\ref{SnowPlowSoln}; dashed).}
\label{FigIsothermalRwithT}
\end{figure}

Fig. \ref{FigIsothermalRwithT} shows the size of the evacuated region
as a function of simulation time for a run with $\lambda = 4$, and the exact
solution for a shell moving in an isothermal background 
(Eq.~\ref{SnowPlowSoln}) with $C$ chosen to match the position of the shell
at $t = 0.1$ Gyr.  The size of the evacuated region is defined by the position
of the gas particle closest to the black hole.  The agreement is very good
with only slight deviation at the latest times. For the model employed, 
the potential is only isothermal near the origin, so when the shell 
expands sufficiently, the potential shallows and the shell should move 
faster than the prediction.  This is indeed seen at late time in 
Fig.~\ref{FigIsothermalRwithT}.

\vspace{-0.3cm}
\subsubsection{Fixed $R_{acc}$}

For the galaxy merger simulations, we apply the force inside a fixed
$R_{acc}$ throughout the simulation.  In this section, we run a
similar set of tests as in the previous subsection but we hold
$R_{acc}$ fixed.  When the number of particles inside $R_{acc}$
becomes small, however, the feedback force exerted on individual
particles becomes spuriously large.  We therefore impose an additional
condition of minimum $N$ on the feedback.  For the tests in this
subsection, the feedback is applied to those particles inside
$R_{acc}$, or to the innermost $100$ gas particles if there are fewer
than this inside $R_{acc}$.  For the simulations in the main paper,
however, there were always enough particles inside the accretion and
feedback region to avoid the need for such a lower bound on $N$.

Our first test uses a constant $L = 4 L_c$, and holds $R_{acc}$ fixed.  
We use a King model as in the previous section, but with slightly
different parameters to connect more closely to the our fiducial 
simulation: $\sigma = 160 \rm{km} \rm{s}^{-1}$, 
$\Psi/\sigma^2 = 12$ and a total mass of $10^{12} M_{\sun}$.  
We tested this model for three different sizes of the accretion and 
feedback region: $R_{acc} = 0.7, 1.4$ and $2.8$ kpc.  The smallest 
region has initially $N \sim 500$.  Note that the values of $R_{acc}$ 
used here are larger than those used in our galaxy merger simulations 
in the main text.  These values of $R_{acc}$ were necessary to ensure 
that $R_{acc}$ contains a reasonable number of particles.  In the 
galaxy merger calculations, the overall larger number of particles in 
the simulation and the high gas density in the central regions imply 
that smaller values of $R_{acc}$ can be reliably used.  They are also 
more physical, as we argued in \S \ref{sectionParam}.

\begin{figure}
\includegraphics[width=84mm]{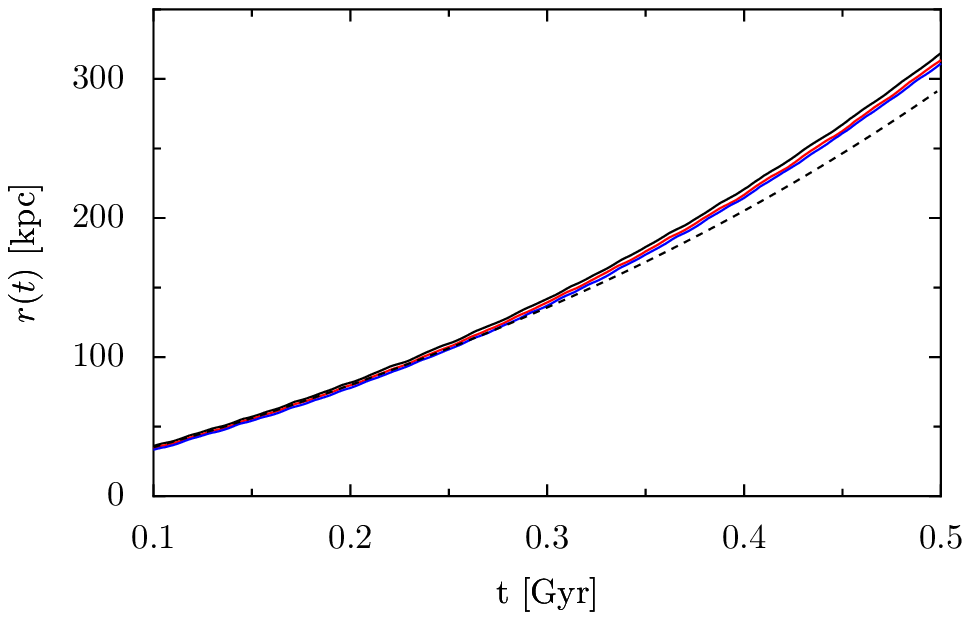}
\caption{Radius of the swept up shell for the isothermal sphere test
  with $\lambda = 4$ and fixed $R_{acc}$: 0.7 (black), 1.4 (red), and
  2.8 kpc (blue).  To avoid numerical problems, the feedback was
  always applied to at least $N \sim 100$ particles.  The numerical
  results agree well with the dashed curve, which shows a numerical
  integration of the analytic equation for the shell radius
  (eq.~\ref{SnowPlowEqn}).}
\label{FigureShellFixed}
\end{figure}

Fig.~\ref{FigureShellFixed} shows the position of the shell of swept
up material for the three runs with $R_{acc} = 0.7, 1.4$ and $2.8$ kpc
in black, red and blue respectively.  Initially, all the gas inside
$R_{acc}$ experiences the extra force.  As the region becomes more
evacuated and the number of particles inside $R_{acc}$ drops, we
transition to applying the force to the $N = 100$ particles closest to
the BH.  The evolution of the shells in this case is quite similar to
the evolution in the last section.  The model used in this section is
smaller in size and so the shell expands past the isothermal core of
the King model earlier. As a result, it begins to accelerate outward
sooner.  However, the tests with different $R_{acc}$ have essentially
identical evolution.

\begin{figure}
\includegraphics[width=84mm]{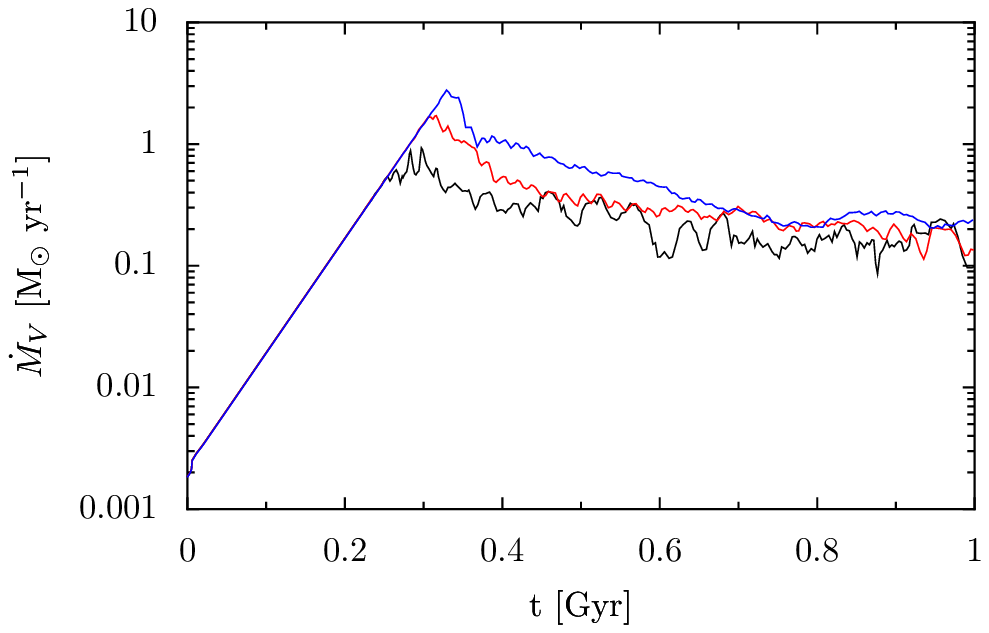}
\caption{The black hole accretion rate for isothermal sphere
  simulations in which the full black hole accretion and feedback
  model are used ($\alpha = 0.1$ and $\tau = 1$).  Three different
  values of $R_{acc}$ are shown: $R_{acc} = 0.7$ (black), 1.4 (red),
  and 2.8 kpc (blue).  All three agree well with each other.}
\label{FigureShellFull}
\end{figure}

Finally, we run a test in which we determine the luminosity from the
accretion rate as in Eq.~(\ref{momdepeqn}), and increase the BH mass 
in time accordingly.  This test thus employs the full feedback and 
accretion model of our galaxy merger simulations.  We use the same 
$\sigma = 160$ km s$^{-1}$ King model, and took $\alpha = 0.1$ and 
$\tau = 1$ for the feedback parameters.  The initial mass of the black 
hole was $M_{BH,i} = 10^5 M_{\sun}$.

Fig.~\ref{FigureShellFull} shows the accretion history of the BH for
the runs with $R_{acc} = 0.7, 1.4$, and 2.8 kpc.  In each test, the
feedback is initially Eddington limited and it is not until about $t =
0.3$ Gyr that the luminosity approaches that required to evacuate the
gas out of $R_{acc}$.  At this point, the gas begins to move out of
$R_{acc}$ and form a shell of material at $R \sim R_{acc}$.  This
shell then remains fairly steady as the accretion rate self-regulates
around the critical luminosity.  As the three values of $R_{acc}$ are
all inside the isothermal core of the King model, the critical
luminosities (eq. \ref{IsoThermCritLEqn}) are the same, and we would
thus expect the accretion rate to self-adjust to the same value at
late times.  This is indeed borne out in the simulations shown in
Fig.~\ref{FigureShellFull}.  Of these three runs, only the
calculation with $R_{acc} = 0.7$ kpc spends a significant amount of
time with fewer than 100 particles inside $R_{acc}$. Despite the large
change in the size of the feedback region, Fig.~\ref{FigureShellFull} 
shows that the evolution of the gas is quite similar.  The black hole 
masses for these three runs differ by only a factor of $\sim 2$ at 
the end of the simulation.

\end{document}